\documentclass[amsmath,amssymb,aps,superscriptaddress,twocolumn,floatfix]{revtex4-2}

\usepackage{graphicx}
\usepackage{caption}
\DeclareCaptionLabelSeparator{dot}{. }
\makeatletter
\def\justified{
	\let\\\@normalcr
	\@rightskip\z@skip \rightskip\@rightskip
	\leftskip\z@skip
	\parindent 0em\relax
	\setlength{\parfillskip}{0pt plus 1fil}}
\DeclareCaptionJustification{justified}{\justified}
\usepackage{subcaption}
\usepackage{amsmath}
\usepackage{braket}
\usepackage{units}
\usepackage{ragged2e}
\usepackage[colorlinks,urlcolor=blue ,citecolor=blue ,linkcolor=blue ]{hyperref}
\usepackage{xcolor}
\usepackage{nicefrac} % for \nicefrac macro
\usepackage{lipsum}
\usepackage{nameref}
\usepackage{hyperref}
\usepackage{textcomp} %for \textmu
\usepackage{urwchancal}
\usepackage{upgreek} %for \upmu
\usepackage{amsmath} % for \multiline
\usepackage{here}
\usepackage{siunitx}

\newcommand{\Tcrit}{T_\text{c} }

\newcommand{\as}{a_\mathrm{s}}
\newcommand{\ad}{a_\mathrm{d}}

\newcommand{\asbg}{$a_\mathrm{s}^\mathrm{bg}$}
\newcommand{\asLMS}{$a_\mathrm{s}^\mathrm{LMS}$}
\newcommand{\sigmaBar}{$\bar{\sigma}$}

\begin{document}

\title{Accurate determination of the scattering length of erbium atoms}

\author{A.\,Patscheider}
\affiliation{Institut f\"ur Experimentalphysik, Universit\"at Innsbruck, Technikerstra{\ss}e 25, 6020 Innsbruck, Austria}

\author{L.\,Chomaz}
\altaffiliation[Present address:]{ Physikalisches Institut, University of Heidelberg, 69120 Heidelberg, Germany.}
\affiliation{Institut f\"ur Experimentalphysik, Universit\"at Innsbruck, Technikerstra{\ss}e 25, 6020 Innsbruck, Austria}

\author{G.\,Natale}
\affiliation{Institut f\"ur Experimentalphysik, Universit\"at Innsbruck, Technikerstra{\ss}e 25, 6020 Innsbruck, Austria}

\author{D.\,Petter}
\altaffiliation[Present address:]{ Optical Materials Engineering Laboratory, Department of Mechanical and Process Engineering, ETH Zurich, 8092 Zurich, Switzerland.}
\affiliation{Institut f\"ur Experimentalphysik, Universit\"at Innsbruck, Technikerstra{\ss}e 25, 6020 Innsbruck, Austria}

\author{M.\,J.\,Mark}
\affiliation{Institut f\"ur Experimentalphysik, Universit\"at Innsbruck, Technikerstra{\ss}e 25, 6020 Innsbruck, Austria}
\affiliation{Institut f\"ur Quantenoptik und Quanteninformation, \"Osterreichische Akademie der Wissenschaften, Technikerstra{\ss}e 21a, 6020 Innsbruck, Austria}

\author{S.\,Baier}
\affiliation{Institut f\"ur Experimentalphysik, Universit\"at Innsbruck, Technikerstra{\ss}e 25, 6020 Innsbruck, Austria}

\author{B.\,Yang}
\altaffiliation[Present address:]{ Southern University of Science and Technology, Shenzhen 518055, China}
\affiliation{Institut f\"ur Experimentalphysik, Universit\"at Innsbruck, Technikerstra{\ss}e 25, 6020 Innsbruck, Austria}

\author{R.\,R.\,W.\,Wang}
\affiliation{JILA and Department of Physics, University of Colorado, Boulder, Colorado 80309, USA}

\author{J.\,L.\,Bohn}
\affiliation{JILA and Department of Physics, University of Colorado, Boulder, Colorado 80309, USA}

\author{F.\,Ferlaino}
\affiliation{Institut f\"ur Experimentalphysik, Universit\"at Innsbruck, Technikerstra{\ss}e 25, 6020 Innsbruck, Austria}
\affiliation{Institut f\"ur Quantenoptik und Quanteninformation, \"Osterreichische Akademie der Wissenschaften, Technikerstra{\ss}e 21a, 6020 Innsbruck, Austria}

\begin{abstract}
    An accurate knowledge of the scattering length is fundamental in ultracold quantum gas experiments and essential for the characterisation of the system as well as for a meaningful comparison to theoretical models. Here, we perform a careful characterisation of the s-wave scattering length $\as$ for the four highest-abundance isotopes of erbium, in the magnetic field range from \SI{0}{G} to \SI{5}{G}. We report on cross-dimensional thermalization measurements and apply the Enskog equations of change to numerically simulate the thermalization process and to analytically extract an expression for the so-called number of collisions per re-thermalization (NCPR) to obtain $\as$ from our experimental data. We benchmark the applied cross-dimensional thermalization technique with the experimentally more demanding lattice modulation spectroscopy and find good agreement for our parameter regime. Our experiments are compatible with a dependence of the NCPR with $\as$, as theoretically expected in the case of strongly dipolar gases. Surprisingly, we experimentally observe a dependency of the NCPR on the density, which might arise due to deviations from an ideal harmonic trapping configuration. Finally, we apply a model for the dependency of the background scattering length with the isotope mass, allowing to estimate the number of bound states of erbium.
\end{abstract}

\date{\today}

\maketitle

\section{Introduction}

The high degree of environmental isolation and the high control over the large parameter-space of ultracold quantum gases are key for their success~\cite{Zwerger:2008}. One of the most decisive properties in determining the many-body phases of a quantum gas is the interaction force between atoms. Among neutral particles, it can be isotropic and short-range, as in alkali atoms, and/or anisotropic and long-range. Open-shell lanthanides, such as erbium (Er) and dysprosium (Dy), have both interactions in place~\cite{Norcia2021nof}. Their strong magnetic character is reflected in a large dipole-dipole interaction (DDI), while the contact potential is governed by the well-known scattering length, whose value $\as$, as in alkali atoms, can be largely controlled by so-called Fano-Feshbach resonances~\cite{Chin2010fri,Frisch2014qci,Maier2015eoc}. 

Although the concept of the scattering length itself is well known by now, theoretical challenges to calculate $\as$ depend on the atomic species of interest. For lanthanides, predicting $\as$ remains a major challenge of quantum chemistry and microscopic scattering theories~\cite{Kotochigova2014}. The complexity of describing such atoms has several reasons: the multiple valence electrons, the strongly anisotropic orbital shells, the strong coupling between core and valence electrons, and the relativistic contributions, also made important by the large atomic mass. To date, there are still no ab-initio models with the capacity for quantitative predictions, although many general properties of the interaction potentials (e.\,g.\,Born-Oppenheimer potentials) have been studied and understood~\cite{Petrov2012aif}. 

Yet, knowledge of the scattering length remains of prime importance since it is an essential regulator of few- and many-body quantum phenomena. For instance, the fascinating supersolid state, recently discovered in Dy~\cite{Boettcher:2019,Chomaz2019lla,Tanzi2019ooa} and Er~\cite{Chomaz2019lla}, lives in a narrow range of only a few \si{\bohr} (\si{\bohr} is the Bohr radius), or the functional forms of beyond-mean-field corrections, which are still under discussion~\cite{Lima2012bmf,Chomaz2018oor,Petter2019ptr,Boettcher2019ddq}, depend on $\as$ in a subtle way.
In the absence of complete microscopic and ab-initio potential models, the study of $\as$ in lanthanides therefore relies on experimental investigations and empirical models.  

Several different experimental methods have been applied in previous works to extract $\as$ for Er and Dy. These include spectroscopy of the molecular binding energy close to a broad Fano-Feshbach resonance~\cite{Maier2015,Lucioni2018}, the anisotropic expansion of a thermal gas~\cite{Tang2016}, and the cross-dimensional thermalization technique~\cite{Jin2014,Aikawa2014ard,Bohn2015,Tang:2015}. Furthermore, for the $^{166}$Er isotope, $\as$ has been determined with high accuracy based on a measurement of the particle-hole excitation gap in the Mott insulator regime via lattice modulation spectroscopy~\cite{Kollath:2006,Baier2016ebh}.
These techniques did not always provide consistent values, opening up a number of fundamental questions, e.\,g.\,from the validity of the additivity of the interaction pseudo-potentials~\cite{Yi2001tco,Ronen2006dbe,Bortolotti2006sli,Oldziejewski2016pos} to the appropriateness of the Lee-Huang-Yang form for beyond-mean field effects~\cite{Chomaz2018oor,FerrierBarbut2018smo,Petter2019ptr,Boettcher2019ddq}.

In this work, we extensively study the scattering length of the four most abundant bosonic isotopes of erbium ($^{164}$Er, $^{166}$Er, $^{168}$Er, and $^{170}$Er) and its magnetic-field dependence. For each isotope, we perform high-resolution Fano-Feshbach spectroscopy in the low magnetic-field region ($0$ to $5$~G) and identify previously unreported scattering resonances. In this range, we then accurately determine the erbium scattering length, $\as$, by developing a model based on the Enskog equations to extract $\as$ from cross-dimensional-thermalization experiments. We benchmark our results with the ones obtained from high-precision lattice-modulation spectroscopy, which has been previously developed for $^{166}$Er~\cite{Baier2016ebh,Chomaz2016qfd} and here expanded to $^{168}$Er.
%Experimentally, we use the technique of cross-dimensional thermalization on a thermal sample and benchmark our results with lattice modulation spectroscopy. Moreover, we exploit the Enskog equations of change to extract an analytic expression for the number of collisions per re-thermalization, as a function of $\as$ and the orientation of the magnetic field and also do a full numerical comparison between theory and experiment. 
Finally, from the magnetic-field mapping of $\as$, we extract for each isotope an effective background scattering length \asbg~at zero $B$ field and we discuss the results in the context of the isotope-mass scaling.

\section{Cross-dimensional Thermalization}

The cross-dimensional-thermalization technique is a very powerful method to experimentally determine the scattering length. First successfully applied to alkali atoms~\cite{Monroe1993,Newbury1995sec,Davis1995eco,Hopkins2000moe}, this technique has proved to be very general and, more recently, has been used for more complex atomic species, such as chromium~\cite{Schmidt2003}, specific isotopes of erbium~\cite{Aikawa2014ard} and dysprosium~\cite{Tang:2015}, and molecular systems~\cite{Valtolina2020deo,Li2021tof}.

Starting from a cold thermal cloud, the basic idea of the cross-dimensional thermalization method is to excite the system by increasing the potential energy along one spatial dimension of the atomic cloud and to measure the characteristic time $\tau$ that the system needs to re-thermalize in an orthogonal directions~\cite{Monroe1993}. In the regime of small excitations, for an atomic cloud at a temperature $T$ and a total atom number $N$, the characteristic time is related to the total scattering cross section $\bar{\sigma}$ by
\begin{equation}
\label{Eq:TauGeneral}
    \tau = \frac{\alpha}{\Bar{n}\Bar{\sigma}v_\text{r}},
\end{equation}
where $\bar{n}$ is the mean number density
\begin{equation}
\label{Eq:meanDensity}
    \Bar{n} = \frac{N \Bar{\omega}^3}{\sqrt{8}} \left( \frac{m}{2\pi k_B T} \right)^{3/2}
\end{equation}
and $v_\text{r}$ the mean relative velocity for two colliding atoms 
\begin{equation}
\label{Eq:relativeVelocity}
    v_\text{r} = \sqrt{\frac{16 k_B T}{\pi m}}
\end{equation}
Here, $\Bar{\omega}$ is the geometric mean of the harmonic trapping frequencies, $m$ is the atomic mass, and $k_B$ is the Boltzmann constant.
Because multiple collisions, not all contributing equally to re-thermalization, are occurring during the thermalization process, the parameter $\alpha$ can be interpreted as a re-scaling of \sigmaBar~and therefore as a number of collisions per re-thermalization (NCPR). Experimentally, the knowledge of $\alpha$ is fundamental for the extraction of the total scattering cross section.

%and $\alpha$ was determined from analytic calculations and Monte-Carlo simulations~\cite{Monroe1993,DeMarco1999},

Equation~\ref{Eq:TauGeneral} has two unknown parameters: $\as$ and $\alpha$. In contrast to alkali atoms, where the scattering is isotropic, the situation is more complex for dipolar atoms such as Er and Dy~\cite{Jin2014,Bohn2015}. Here, the total cross section for bosons is not only given by the contact scattering length $\as$, but an additional contribution from the non-isotropic DDI, which for two atoms at a distance $r$ and polarized by an external magnetic field $\mathbf{B}$, reads as
\begin{equation}
    V_\text{dd}(r,\theta) = \frac{\mu_0 \mu^2}{4\pi}\frac{1-3\cos^2 \theta}{|\mathbf{r}|^3}.
\end{equation}
Here, $\mu_0$ is the magnetic permeability, $\mu$ is the magnetic dipole moment, and $\theta$ the angle between $\mathbf{B}$ and $\mathbf{r}$.
Taking an angular average of the total cross section leads to
\begin{equation}
    \label{Eq:sigma_bar}
    \bar{\sigma} = 8\pi a_s^2 + \frac{32\pi}{45}a_\text{d}^2,
\end{equation}
where $\ad = \frac{m \mu_0 \mu^2}{8\pi\hbar^2}$ is the dipolar length ($\ad = \SI{98.2}{\bohr}$ for $^{166}$Er), with $\hbar$ being the reduced Planck constant. Finally, we can rewrite Eq.~\ref{Eq:TauGeneral} as 
\begin{equation}
\label{Eq:TauSpecific}
    \tau = \frac{\alpha}{\Bar{n}\Bar{\sigma}v_\text{r}} = \frac{\alpha}{\frac{4Nm\Bar{\omega}}{\pi k_\text{B} T_0} (a_\text{s}^2 + \frac{4}{45}\ad^2)}.
\end{equation}
The interplay between the isotropic scattering length and the anisotropic dipolar cross section leads to a dependence of $\alpha$ on both, the dipole orientation $\theta$ and $\as$~\cite{Wang2021}. In the limit of weak excitation, an analytic form of $\alpha (\as,\theta)$ can be found based on the Enskog equations; see later discussion.

\section{Experimental procedure}

\begin{figure}
    \centering
    \includegraphics[width = 0.48\textwidth]{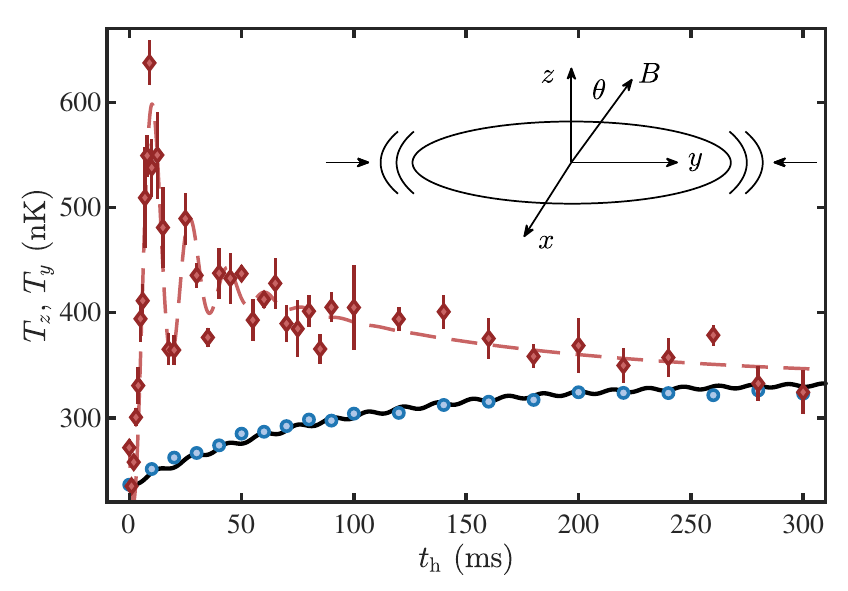}
    \caption{Effective temperatures $T_z$ (blue circles) and $T_y$ (red diamonds) after the increase of the trapping potential along the weakest trapping direction $y$. The measurement was performed at \SI{1}{G} and $\theta = \SI{0}{\degree}$ for the $^{166}$Er isotope. The red dashed line represents a guide to the eye. The black solid line denotes the results of the Enskog simulations for this specific dataset. The errorbars denote the standard error for $3$ repetitions. The inset shows a schematic representation of our experimental system.}
    \label{fig:Figure_CT_and_System}
\end{figure}

In our experiment, we produce a spin-polarized thermal cloud of Er atoms in the lowest Zeeman sublevel, similarly to Ref.\,\cite{Aikawa2012bec}. In brief, after cooling and trapping the Er atomic ensemble in a narrow-line magneto-optical trap~\cite{Frisch2012nlm}, we transfer the atoms into a crossed optical dipole trap (cODT). Here, we first further cool the atoms via standard evaporative cooling, and then tighten the trapping confinement to avoid atom loss due to residual evaporation. Simultaneously, we ramp $\mathbf{B}$ to the desired value. At this stage we typically reach a temperature of $T = \SI{250}{nK}-\SI{300}{nK}$ with $N \approx \num{1e5}$. The exact numbers depend on the isotope choice and the individual set of measurement. The typical final trap frequencies are $(\omega_x, \omega_y, \omega_z) = 2\pi\times (65(1), 19(1), 300(2))\,\si{Hz}$. For all sets of measurements the critical temperature for the onset of Bose-Einstein condensation $\Tcrit$ lies between \SI{150}{nK} and \SI{200}{nK}, such that $T \gtrsim 1.5 \times \Tcrit$. The orientation of the magnetic dipoles is controlled by the direction of the polarizing $\mathbf{B}$ and is represented by the angle $\theta$ between $\mathbf{B}$ and the vertical direction $z$, defined by gravity; see inset Fig.\,\ref{fig:Figure_CT_and_System}.

After preparing the thermal sample, we perform cross-dimensional thermalization experiments~\cite{Aikawa2014ard}. In particular, we excite the cloud along the $y$ direction and probe the thermalization dynamics in the $z$ direction. Our excitation scheme relies on a rapid increase in power of one trapping beam, leading to a \SI{60}{\%} increase of the trapping frequency, while leaving the other two directions mostly unaffected. We extract the effective temperature $T_z$ ($T_y$) for a variable in-trap hold time $t_\mathrm{h}$ from the width of the momentum distribution $\sigma_{z}(t_\mathrm{h})$ ($\sigma_{y}(t_\mathrm{h})$) after a time of flight of $t_\mathrm{ToF}=\SI{25}{ms}$ ($\SI{20}{ms}$). This scheme, illustrated in the inset of Fig.\,\ref{fig:Figure_CT_and_System}, leads to an out-of-equilibrium cloud with an effective temperature increase along $y$ from about \SI{300}{nK} to \SI{600}{nK}. 

Figure~\ref{fig:Figure_CT_and_System} shows $T_z$ and $T_y$ as a function of $t_\mathrm{h}$ at $B =  \SI{1}{G}$. As we excite the system along $y$, we observe the expected rapid increase of $T_y$. After reaching a maximum effective temperature, $T_y$ starts to decay, and simultaneously $T_z$ increases, both reaching the same equilibrium temperature, thus showing thermalization dynamics. We observe oscillations in $T_y$, which we attribute to a breathing mode that gets induced by the excitation. We observe an exponential-type growth of the form 
\begin{equation}
\label{eq:ExponentialThermalization}
    T_z(t) = T_\mathrm{f} (1-\Delta T  e^{-t/\tau}).    
\end{equation}
Here, $T_\mathrm{f}$ denotes the final temperature and $\Delta T$ denotes the temperature increase due to the added energy. However, using this simple fit we can not directly extract $\as$ as additional knowledge on $\alpha (\as,\theta)$ is needed (see Eq.\,\ref{Eq:TauSpecific}).

\begin{figure}
    \centering
    \includegraphics[width = 0.48\textwidth]{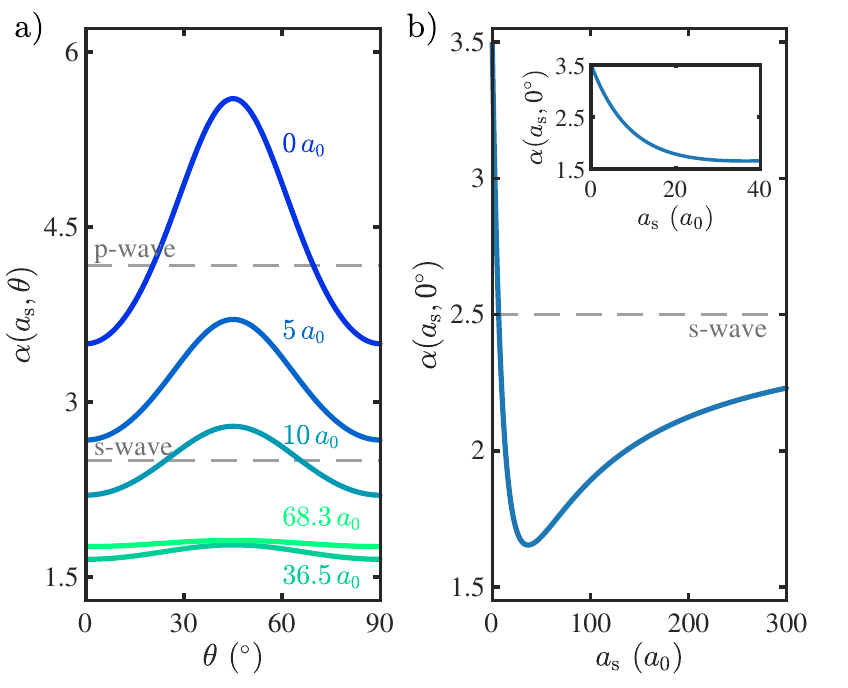}
    \caption{(a) Dependency of $\alpha$ on $\theta$ and $\as$~for $a_\mathrm{s} = \SI{0}{\bohr}$, \SI{5}{\bohr}, \SI{10}{\bohr}, \SI{36.5}{\bohr}, and \SI{68.3}{\bohr}. These values are chosen such that the angle dependence at small $\as$ becomes visible. Note that, at \SI{68.3}{\bohr} ($\as$ at \SI{1}{G}, see later measurements) the variation of $\alpha$ with $\theta$ is strongly suppressed. (b) $\alpha$ vs. $\as$~for $\theta = \SI{0}{\degree}$. The inset shows an enlargement of the region for $\as$ between \SI{0}{\bohr} and \SI{40}{\bohr}. The grey dashed lines show the values of $\alpha$ for s-wave and p-wave scattering, respectively. }
    \label{fig:alpha_vs_theta}
\end{figure}

\section{Theoretical estimate of $\alpha (\as, \theta)$}

To compute $\alpha (\as, \theta)$, we utilize 
% a theoretical model based on solutions from 
the Enskog equations of change~\cite{Reif1965}: a coupled set of differential equations derived in closed-form for dipolar gases, by linearization of the Boltzmann equation, and the assertion of a Gaussian phase-space distribution~\cite{Wang2020}. 
% The Enskog equations have been used to derive 
These equations permit an analytic derivation of $\alpha (\as,\theta)$ in the limit of short-times and small excitations~\cite{Wang2021}. For the current experiment with excitation along $y$ and thermalization measured along $z$, the NCPR is described by a simple analytic formula, which reads
\begin{align}
    \alpha(\as,\theta) = \frac{ 14 \left(45 a_s^2 + 4 \ad^2\right) }{ 252 a_s^2 + 96 a_s \ad + (3 \cos (4 \theta ) + 13) \ad^2 }. \label{eq:analytic_alpha}
\end{align}
The quantity $\alpha (\as,\theta)$ exhibits an anisotropic character via its angle dependence, as already observed for dipolar fermionic atoms~\cite{Aikawa2014ard} and molecules~\cite{Li2021tof}.

\begin{figure*}
    \centering
    \includegraphics[scale =1]{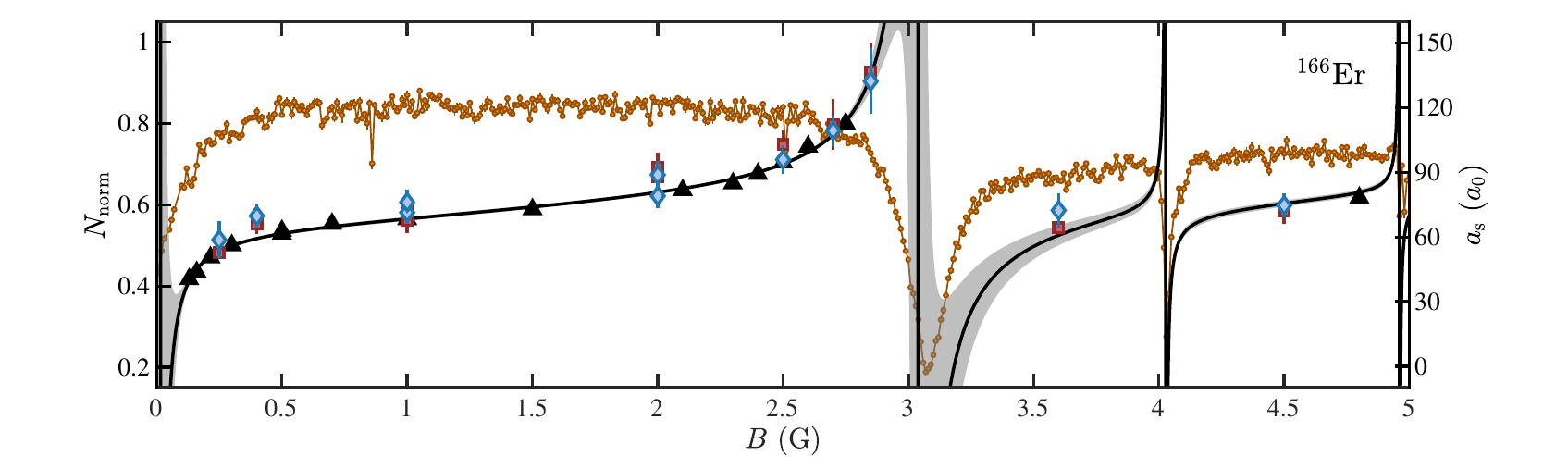}
    \caption{Normalized atom number (orange circles) and $\as$~extracted from cross-dimensional thermalization measurements using both, the Enskog equations (red squares) and the analytic formula of Eq.~\ref{eq:analytic_alpha} (blue diamonds) are shown for $^{166}$Er. Additionally \asLMS~(black triangles) obtained from lattice modulation spectroscopy measurements are given. The solid black lines represent a fit of Eq.\,\ref{Eq:a_background_fit} to \asLMS. Error bars and the shaded area of the fitting results denote the standard error.}
    \label{fig:as_166}
\end{figure*}

Figure~\ref{fig:alpha_vs_theta} shows $\alpha (\as,\theta)$ as a function of $\theta$ (a) and $\as$ (b), for our experimental configuration of a pancake shaped trap. Figure~\ref{fig:alpha_vs_theta}(a) shows that the anisotropic character of $\alpha (\as,\theta)$ competes with the contact one. Indeed, while for small $\as$~($\lesssim \SI{10}{a_0}$), $\alpha (\as,\theta)$ exhibits a pronounced angle-dependence with a maximum at \SI{45}{\degree}, for increasing $\as$ such behavior progressively washes out. For $\as \approx \SI{70}{\bohr}$, the thermalization behavior becomes basically independent of $\theta$, however $\alpha (\as,\theta)$ acquires a number below the one expected for purely contact interacting s-wave collisions. This suggests faster thermalization for dipolar particles, arising from a more efficient diversion of velocities of the scattering constituents. In the experiment, we only measure re-thermalization for  relatively large values of $\as \gtrsim \SI{30}{\bohr}$, and therefore we are not sensitive to the angle dependence of $\alpha (\as,\theta)$. In the course of this work, we will thus focus on the case $\theta = \SI{0}{\degree}$, simplifying Eq.\,\ref{eq:analytic_alpha} to 
\begin{equation}
    \alpha(\as,\theta = \SI{0}{\degree}) = \frac{ 14 \left(45 a_s^2 + 4 \ad^2\right) }{ 252 a_s^2 + 96 a_s \ad + 16 \ad^2 }.
    \label{eq:analytic_alpha_theta_0}
\end{equation}

As shown in Figure~\ref{fig:alpha_vs_theta}(b), after an initial decrease, $\alpha (\as,\SI{0}{\degree})$ increases for $\as \gtrsim \SI{36.7}{\bohr}$ -- and thus the thermalization loses efficiency -- moving to the regime of contact dominated interaction, eventually reaching the $\alpha (\as,\SI{0}{\degree}) = \num{2.5}$ limit of non-magnetic atoms~\cite{DeMarco1999,Jin2014}. We note that by setting $\theta = \SI{0}{\degree}$ and $\ad/\as \approx \num{2.7}$, the NCPR is minimized with value $\alpha \approx 1.65$, indicating highly efficient collisional thermalization. This is directly attributed to the innate anisotropic differential cross-section in dipolar bosons \cite{Jin2014}.    

\section{Mapping of $\as$ as a function of $\mathbf{B}$ for $^{166}\text{Er}$}

Before taking cross-dimensional thermalization measurements for $^{166}$Er, we perform a high resolution scan of the atom number as a function of the magnetic field in order to record the spectrum of Fano-Feshbach resonances, which we know to be exceptionally dense~\cite{Frisch2014qci,Maier2015eoc}. We record the Fano-Feshbach spectra in a magnetic field region from \SI{0}{G} to \SI{5}{G}; see Fig.\,\ref{fig:as_166} and Ref.\,\cite{supmat}. In all the measurements the magnetic field is oriented along $z$.

We then perform thermalization measurements at values of the magnetic field, where the system is not dominated by resonant atom loss. For each thermalization curve, we extract $\as$ using two different approaches, one numerical and one semi-analytical. The first, constitutes a direct fit of the full Enskog solutions to the experimental data, leaving $\as$ as a float parameter of the theory; see~\cite{supmat} for more details. The second method, is based on the exponential growth rate $\tau$, from Eq.\,\ref{Eq:TauGeneral} using the analytic expressin for $\alpha (\as,\SI{0}{\degree})$ in Eq.\,\ref{eq:analytic_alpha_theta_0}. For the latter, since $\as$ is unknown a priori, we use an iterative approach to determine $\alpha (\as,\SI{0}{\degree})$ starting from $\alpha (\as,\SI{0}{\degree}) = \num{1.7}$. We use the calculated $\as$ and the analytic formula (see Eq.~\ref{eq:analytic_alpha_theta_0}) to obtain a new value for $\alpha (\as,\SI{0}{\degree})$. We stop the iteration once the relative change of $\alpha (\as,\SI{0}{\degree})$ is $\leq \num{1e-7}$.

Figure~\ref{fig:as_166} summarizes $\as$ for $^{166}$Er in the region from \SI{0}{G} to \SI{5}{G}. In the studied $B$-field regime, the scattering behavior is essentially dominated by a broad resonance at \SI{3}{G} and a second one around $B=\SI{0}{G}$. The $\as$ extracted from the Enskog model and the semi-analytic one are in very good agreement with each other, reflecting the strength of the analytic formula of Eq.\,\ref{eq:analytic_alpha_theta_0}. 

\section{Benchmarking with lattice spectroscopy}

To evaluate the robustness of our approach to extract $\as$, we benchmark our cross-dimensional thermalization results with the one obtained using an alternative technique based on lattice modulation spectroscopy (LMS). Such a technique, which we have developed in the past for $^{166}$Er~\cite{Baier2016ebh,Chomaz2016qfd} and $^{167}$Er~\cite{Baier2018roa}, is based on the measurement of the on-site interaction - related to $\as$ - of a lattice-confined dipolar gas in a Mott insulator state. The LMS is able to provide accurate values of \asLMS, but at the price of being experimentally more involved due to its requirements of an optical lattice together with a clean degenerate sample. Here we compare the values of $a_s$ obtained with cross-dimensional thermalization on a low-density thermal sample, with \asLMS~obtained from the lattice modulation spectroscopy obtained in Ref.\,\cite{Chomaz2016qfd}.
In brief we extract \asLMS~as follows. We prepare an ultracold sample of $^{166}$Er atoms in a three-dimensional optical lattice, created by two retro-reflected laser beams at \SI{532}{nm} in the horizontal plane and by one retro-reflected laser beam at \SI{1064}{nm} along the vertical $z$ direction, defined by gravity. The final lattice depth along the three directions is  $(s_x, s_y, s_z) = (20, 20, 100)$, in units of $E_{\text{rec}} = \SI{4.2}{kHz}
~(\SI{1.05}{kHz})$ for \SI{532}{nm} (\SI{1064}{nm}). The uncertainty on $s_x$, $s_y$, and $s_z$ is about $\SI{5}{\%}$. In such a deep lattice, the atoms are in the Mott insulator phase~\cite{Baier2016ebh}.

%but at the price of being experimentally heavy, making extensive measurements as a function of the magnetic field rather demanding.

We then create particle-hole-excitations by sinusoidally modulating the power of the horizontal lattice beams for \SI{90}{ms} with a peak-to-peak amplitude of about \SI{30}{\%} and measure the recovered BEC fraction after melting of the lattice. At the resonance condition, where the modulation frequency matches the particle-hole excitation gap, we observe a resonant reduction in the BEC fraction~\cite{Greiner2002qpt}. The particle-hole excitation gap is directly given by the on-site interaction $U = U_c + U_{dd}$. Here,  $U_c$ is the contact interaction -- and thus depends on the unknown \asLMS -- while the on-site dipolar interaction, $U_{dd}$, can be accurately calculated. We repeat the measurements at various magnetic-field values and, for each, we extract \asLMS.

In Fig.~\ref{fig:as_166}, we compare \asLMS~with $\as$ extracted from the thermalization measurements. We see an overall very good agreement between the value of $\as$ extracted using the two techniques. This shows that the cross-dimensional thermalization approach combined with the Enskog equations is a very reliable method to extract $\as$, even in the case of complex atoms for which the knowledge of $\alpha (\as,\theta)$ is not a priori given. 

\section{Density dependence}

Our measurements for the $^{166}$Er isotope were performed in a regime of relatively low density ($\bar{n} \leq \SI{0.5e13}{cm^{-3}}$). Interestingly, when applying the same method in a regime of high density, we observe a dependence of the thermalization rate on the density which goes beyond the Enskog approach. For instance, we repeat the cross-dimensional thermalization measurements for $^{166}$Er at $B = \SI{1}{G}$ and variable cloud density, $\bar{n}$. We control the density by either increasing $N$ or by applying a tighter trapping configuration of $(\omega^\mathrm{cyl}_x, \omega^\mathrm{cyl}_y, \omega^\mathrm{cyl}_z) \approx 2\pi\times (300, 19, 300)\,\si{Hz}$ before compression, or both. From the lattice modulation spectroscopy, we have extracted a value $\as=\SI{68.3(7)}{\bohr}$ at $B=\SI{1}{G}$. By fixing this value -- meaning to impose that the scattering length does not depend on density -- and using Eq.\,\ref{Eq:TauGeneral}, we can determine $\alpha (\as,\SI{0}{\degree})$ as a function of $\bar{n}$.

\begin{figure}
    \centering
    \includegraphics[width = 0.48\textwidth]{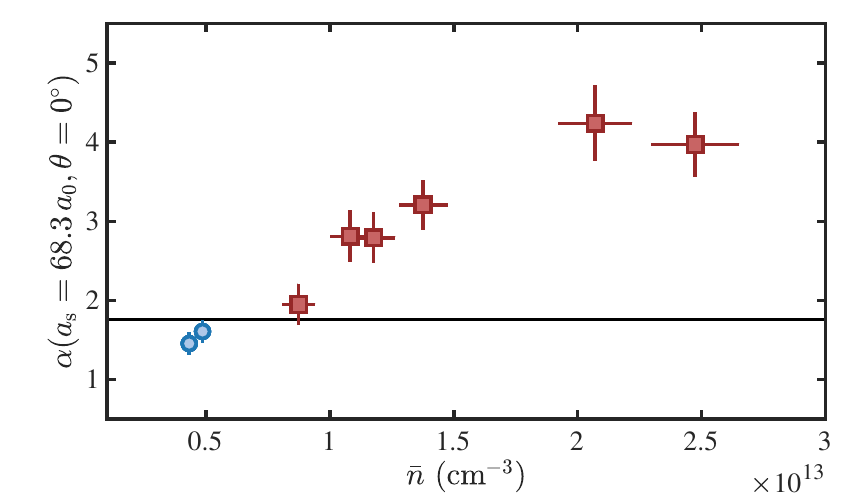}
    \caption{Measurements of $\alpha (\as,\theta)$ as a function of $\bar{n}$. The blue circles correspond to the datasets at \SI{1}{G}, shown in Fig.\,\ref{fig:as_166}. The black solid line marks the value given by the analytic formula in Eq.\,\ref{eq:analytic_alpha_theta_0}. All measurements are performed with $\theta = \SI{0}{\degree}$. Errorbars denote the standard error.}
    \label{fig:alpha_n}
\end{figure}

Figure~\ref{fig:alpha_n} shows $\alpha (\as,\SI{0}{\degree})$ for different values of $\bar{n}$. We find a pronounced dependency on $\bar{n}$, with a rapid increase and an eventual saturation at high densities. Such a behavior is not captured by our theoretical model, which, as reflected in the definition of $\alpha (\as,\SI{0}{\degree})$ in Eq.\,\ref{Eq:TauGeneral}, predicts no density dependence. To the best of our knowledge, such a dependence has not been reported in previous works on cross-dimensional thermalization. Possible explanations root in various causes, either physical or technical nature. Although being above $\Tcrit$, precursors of quantum many-body phenomena might influence the scattering behavior. Exemplary, we tried to explicitly include effects coming from Bose-enhancement into our theoretical framework. This did not have significant influence on the thermalization behavior.

Another possible explanation, based on unavoidable experimental imperfections, roots in deviations from an ideal harmonic trapping condition, leading to a modification of the kinetic energy and the mean density. Such a variation would manifest in an apparent change of $\alpha (\as,\SI{0}{\degree})$; see Eq.\,\ref{Eq:TauGeneral}. Indeed, Eq.\,\ref{Eq:meanDensity} and \ref{Eq:relativeVelocity} are only valid for an ideal harmonic trapping confinement. First Monte-Carlo simulations performed by using a realistic gaussian trapping potential seem to support this assumption~\cite{supmat}. Note that, varying the initial temperature of the atomic cloud and the excitation strength did not show any influence on the observations. We emphasize that, since our measurements to extract $\as$ have been performed at low densities, our method should remain valid.   

%A possible explanation lies in a modification of the kinetic energy as well as the mean density due to deviations from an ideal harmonic trapping condition. Such a variation would manifest in an apparent change of $\alpha (\as,\SI{0}{\degree})$; see Eq.\,\ref{Eq:TauGeneral}. Indeed, Eq.\,\ref{Eq:meanDensity} and \ref{Eq:relativeVelocity} are only valid for an ideal harmonic trapping confinement. First Monte-Carlo simulations performed by using a realistic gaussian trapping potential seem to support this assumption~\cite{supmat}. Note that, varying the initial temperature of the atomic cloud and the excitation strength did not show any influence on the observations. We emphasize that, since our measurements to extract $\as$ have been performed at low densities, our method should remain valid. 

\section{Scattering length for $^{164}\text{Er}$ and $^{170}\text{Er}$}

\begin{figure*}
    \centering
    \includegraphics[scale =1]{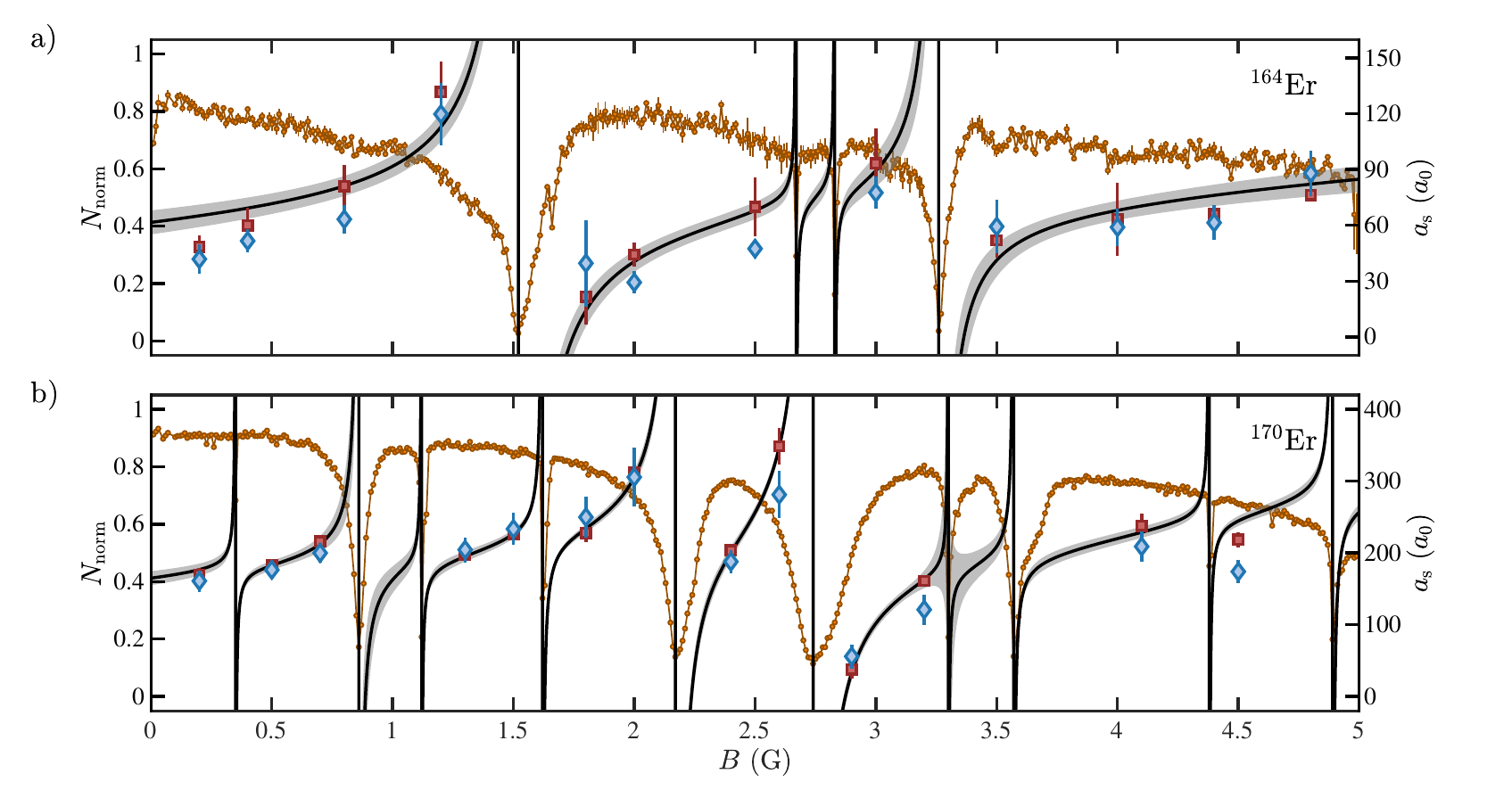}
    \caption{Normalized atom number (orange circles) and $\as$~extracted from cross-dimensional thermalization measurements using both, the Enskog equations (red squares) and the analytic formula of Eq.~\ref{eq:analytic_alpha} (blue diamonds) are shown for (a) $^{164}$Er and (b) $^{170}$Er. The solid black lines represent a fit of Eq.\,\ref{Eq:a_background_fit} to $\as$~obtained using the Enskog equations. Error bars and the shaded area of the fitting results denote the standard error.}
    \label{fig:as_164_170}
\end{figure*}

After the detailed study on $^{166}$Er and the benchmarking of the results with high-precision lattice modulation spectroscopy, we confidently apply our cross-dimensional thermalization approach to other two isotopes, $^{164}$Er and $^{170}$Er. Again we start with a Fano-Feshbach spectroscopy between \SI{0}{G} and 5G to identify the position of the scattering resonances as shown in Fig.\,\ref{fig:as_164_170}. We note that this Fano-Feshbach spectra have not been reported previously. For the cross-dimensional thermalization measurements we follow a similar experimental procedure as described above. From the thermalization curve, we again use both, the full fit of the Enskog equations as well as the iterative approach on $\alpha (\as,\SI{0}{\degree})$ to determine $\as$ from the exponential growth rate $\tau$.

Figure~\ref{fig:as_164_170} shows $\as$ for the isotopes (a) $^{164}$Er and (b) $^{170}$Er. While the scattering behavior for $^{164}$Er is, similarly to $^{166}$Er, dominated by two broad resonances at \SI{1.5}{G} and \SI{3.3}{G}, $^{170}$Er features several narrow overlapping resonances, providing different test scenarios for our cross-dimensional thermalization. Although minor deviations can be observed in the vicinity of Fano-Feshbach resonances, for both isotopes, the extracted $\as$ using the two approaches are once more in good agreement. 

\section{Scaling of background scattering length with mass}

The knowledge on $\as$  as a function of the magnetic-field allows us to extract an effective background scattering length $a_\mathrm{s}^\text{bg}$ for each isotope. The general behavior of $\as$~with $B$ can be described by generalizing the well-known formula~\cite{Jachymski2013}
\begin{equation}
\label{Eq:a_background_fit}
   a_\mathrm{s}(B) = (a_\mathrm{s}^\text{bg} + sB) \times \prod_{i=1}^{N_\text{res}} \left( 1 - \frac{\Delta B_i}{B - B_i} \right),
\end{equation}
to the case of $N_\text{res}$ overlapping resonances of position $B_i$ and width $\Delta B_i$, and allowing for a smooth off-resonant  variation of $\as$ with $B$. We observe that a linear variation of slope $s$ already well reproduces the data with $a_\mathrm{s}(0)$ defined as the effective $a_\mathrm{s}^\text{bg}$. We note that different mechanisms could lead to an off-resonant  variation of $\as$. For instance, the influence of broad Fano-Feshbach resonances, which are not within our measurement range, could lead to a smooth variation of the background behavior, similar to that observed for cesium~\cite{Kraemer2006efe}. Alternatively, the effect could be due to the coupling induced by DDI between the incident scattering channel and Zeeman states that lie higher in energy. As a consequence this results in a perturbation of the molecular potential, whose strength depends on the magnetic field, leading to an increasing value of the van der Waals $C_6$ coefficient~\cite{Deb2001}. %At larger $B$, this coupling becomes weaker and thereby influences $\as$. 

To parametrize $\as$ as a function of $B$, we fit Eq.\,\ref{Eq:a_background_fit} to the measured $\as$ for $^{164}$Er, $^{166}$Er, $^{168}$Er, and $^{170}$Er. For $^{166}$Er and $^{168}$Er, we use the scattering lengths obtained from the lattice modulation spectroscopy, corresponding to our most accurate determination; see solid lines in Fig.\,\ref{fig:as_166} and \cite{supmat}. For $^{164}$Er and $^{170}$Er, we fit Eq.\,\ref{Eq:a_background_fit} to the $\as$ data obtained by applying the Enskog equations to the cross-dimensional thermalization measurements; see solid lines in Fig.\,\ref{fig:as_164_170}. More details on the fitting procedure as well as the complete list of the fit parameters is given in Ref.\,\cite{supmat}. In general, we observe that the fitting function reproduces very well the  behavior of $\as$ for every isotope.

\begin{figure}
    \centering
    \includegraphics[width =0.48\textwidth]{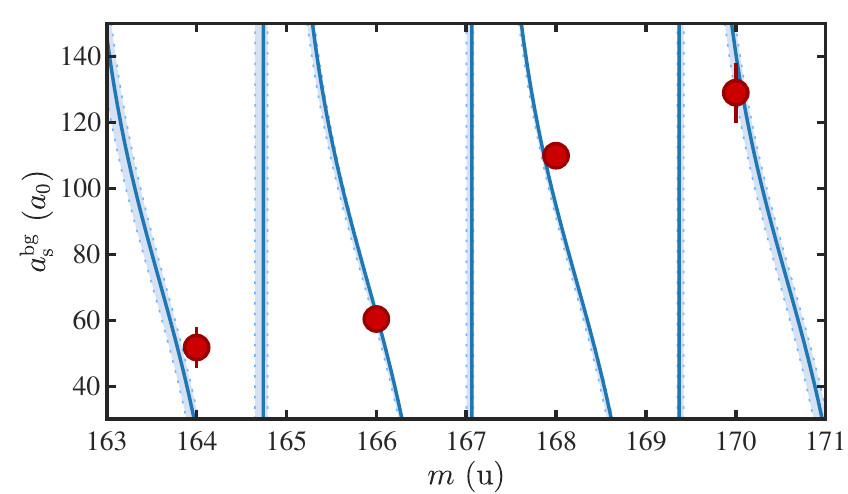}
    \caption{Background scattering length \asbg~for four bosonic isotopes (red circles). The solid line represents the best fit with $\phi = 144(1)$; see text. The shaded area, enclosed by the dotted lines, represents the fitting function for $\phi = 143$ and $\phi = 145$. The errorbars denote the standard error of the fit of Eq.\,\ref{Eq:a_background_fit} to the experimental data.}
    \label{fig:mass_scaling}
\end{figure}

Figure~\ref{fig:mass_scaling} shows the value of \asbg~from the fit as a function of the isotope mass. We observe a monotonic rising of \asbg~with increasing $m$, which might be compatible with different functional forms, including a simple linear increase. Under the assumption that erbium has a similar behavior to ytterbium and cesium, we can use the model for the mass scaling as developed in Ref.\,\cite{Gribakin1993,Kitagawa2008,Borkowski2013sli}. Such a model assumes that $\as$ is only given by the Van der Waals potential $U(r) = -C_6/r^6$, with $C_6$ being the Van der Waals coefficient. This might be a rather severe approximation for magnetic atoms but, in absence of  alternative models, it is interesting to compare the simple mass-scaling approach to erbium.

As introduced in Ref.\,\cite{Gribakin1993}, $\as$~can be written as
\begin{equation}
    \label{Eq:a_s_scaling_full_form}
    a_\mathrm{s} = \bar{a} \left[ 1 - \tan \left(\phi - \frac{\pi}{8}\right) \right], 
\end{equation}
with $\bar{a} = 2^{-3/2} \frac{\Gamma (3/4)}{\Gamma (5/4)} \left( \frac{m C_6}{\hbar^2} \right)^{1/4}$ being the characteristic length and
\begin{equation}
    \phi = \frac{\sqrt{m}}{\hbar}\int_{R_0}^\infty \sqrt{-U(r)}dr.
\end{equation}
Here, $\Gamma(x)$ is the gamma-function and $R_0$ is the classical turning point of $U(r)$. Although the exact shape of $U(r)$ is unknown, Eq.\,\ref{Eq:a_s_scaling_full_form} can be employed to extract a mass-scaling due to the dependence of $\phi \propto \sqrt{m}$~\cite{Kitagawa2008}. Such a scaling is valid, as long as the mass dependent modification of $U(r)$ is negligible. Furthermore, $\phi$ allows for the calculation of the number of bound states $N_\mathrm{B}$ via relation $N_\mathrm{B} = \lfloor \phi/\pi - 5/8 \rfloor$, where $\lfloor~ \rfloor$ denotes the floor integer function. 

We now apply this model to our Er case. Figure~\ref{fig:mass_scaling} shows the fit of Eq.\,\ref{Eq:a_s_scaling_full_form} to the experimental data; see Ref.\,\cite{supmat} for details. We obtain the best agreement for $\phi = 144(1)$, leading to $N_\mathrm{B} = 143(1)$ for $^{168}$Er. Despite the similar $C_6$ coefficient, $N_B$ is approximately a factor of $2$ larger than for ytterbium~\cite{Kitagawa2008}. Note that, $N_\mathrm{B}$ is in agreement with the result obtained when using the same approach but assuming a hard core potential~\cite{supmat}. We would like to emphasize once more that this model does not consider any contribution arising from the DDI. An improved description calls for the development of advanced theoretical models. 

\section{Conclusion}

In conclusion, we report on an accurate study of the scattering length of four different isotopes of erbium.  Our work focuses on the low magnetic field region, which is the range of most interest in current experiments. Our experimental survey combines two different techniques: a high-precision, yet demanding, approach based on the measurement of the onsite interaction in a Mott insulator phase, and another one based on measuring the re-equilibration time in cross-dimensional thermalization experiments. From the latter, we extract the the value of $\as$ by both numerically applying the full Enskog equations and using the analytic formulation for $\alpha (\as,\theta)$. 
All these different approaches, benchmarked one with respect to the others, provide a very consistent measure of the scattering length in the region of interest. These results will be relevant for current experiments and moreover point to a practical manner to extract $\as$ with reduced experimental effort, which can be readily generalized to other magnetic lanthanides. 
%cross-dimensional thermalization measurements allow to extract $\as$ with reduced experimental effort as compared to the lattice modulation spectroscopy. We determine the scattering length $\as$ from thermalization measurements for three bosonic isotopes over a magnetic field range from \SI{0}{G} to \SI{5}{G}. We successfully benchmark our outcomes for one isotope by comparison to lattice modulation spectroscopy results. Surprisingly, we find a so far unobserved dependence of $\alpha (\as,\theta = \SI{0}{\degree})$ on $\bar{n}$, which might arise due to a non-perfect harmonic trapping configuration. Our finding represents an important input for future experimental work on thermalization measurements with dipolar atoms as well as for theoretical models. Finally, we use our results to extract the background scattering length for four isotopes. We further apply a simple theoretical model to the scaling of the latter with isotope mass, allowing to extract the number of bound states for erbium. As the model considers only the van der Waals interaction potential, our work constitutes the basis for advanced theoretical work including contributions from DDI. 

\begin{acknowledgments}
This work is financially supported through an ERC Consolidator Grant (RARE, no.\,681432), a DFG/FWF (FOR 2247/I4317-N36) and a joint-project grant from the FWF (I4426, RSF/Russland 2019). We also acknowledge the Innsbruck Laser Core Facility, financed by the Austrian Federal Ministry of Science, Research and Economy. R.\,R.\,W.\,W.\,and J.\,L.\,B.\,acknowledge that this material is based on work supported by the National Science Foundation under grants numbered PHY-1734006 and PHY-2110327.
\end{acknowledgments}

* Correspondence and requests for materials
should be addressed to Francesca.Ferlaino@uibk.ac.at.

\bibliography{BibScattLengthArticle}

\begin{thebibliography}{54}%
\makeatletter
\providecommand \@ifxundefined [1]{%
 \@ifx{#1\undefined}
}%
\providecommand \@ifnum [1]{%
 \ifnum #1\expandafter \@firstoftwo
 \else \expandafter \@secondoftwo
 \fi
}%
\providecommand \@ifx [1]{%
 \ifx #1\expandafter \@firstoftwo
 \else \expandafter \@secondoftwo
 \fi
}%
\providecommand \natexlab [1]{#1}%
\providecommand \enquote  [1]{``#1''}%
\providecommand \bibnamefont  [1]{#1}%
\providecommand \bibfnamefont [1]{#1}%
\providecommand \citenamefont [1]{#1}%
\providecommand \href@noop [0]{\@secondoftwo}%
\providecommand \href [0]{\begingroup \@sanitize@url \@href}%
\providecommand \@href[1]{\@@startlink{#1}\@@href}%
\providecommand \@@href[1]{\endgroup#1\@@endlink}%
\providecommand \@sanitize@url [0]{\catcode `\\12\catcode `\$12\catcode
  `\&12\catcode `\#12\catcode `\^12\catcode `\_12\catcode `\%12\relax}%
\providecommand \@@startlink[1]{}%
\providecommand \@@endlink[0]{}%
\providecommand \url  [0]{\begingroup\@sanitize@url \@url }%
\providecommand \@url [1]{\endgroup\@href {#1}{\urlprefix }}%
\providecommand \urlprefix  [0]{URL }%
\providecommand \Eprint [0]{\href }%
\providecommand \doibase [0]{https://doi.org/}%
\providecommand \selectlanguage [0]{\@gobble}%
\providecommand \bibinfo  [0]{\@secondoftwo}%
\providecommand \bibfield  [0]{\@secondoftwo}%
\providecommand \translation [1]{[#1]}%
\providecommand \BibitemOpen [0]{}%
\providecommand \bibitemStop [0]{}%
\providecommand \bibitemNoStop [0]{.\EOS\space}%
\providecommand \EOS [0]{\spacefactor3000\relax}%
\providecommand \BibitemShut  [1]{\csname bibitem#1\endcsname}%
\let\auto@bib@innerbib\@empty
%</preamble>
\bibitem [{\citenamefont {Bloch}\ \emph {et~al.}(2008)\citenamefont {Bloch},
  \citenamefont {Dalibard},\ and\ \citenamefont {Zwerger}}]{Zwerger:2008}%
  \BibitemOpen
  \bibfield  {author} {\bibinfo {author} {\bibfnamefont {I.}~\bibnamefont
  {Bloch}}, \bibinfo {author} {\bibfnamefont {J.}~\bibnamefont {Dalibard}},\
  and\ \bibinfo {author} {\bibfnamefont {W.}~\bibnamefont {Zwerger}},\
  }\bibfield  {title} {\bibinfo {title} {Many-body physics with ultracold
  gases},\ }\href {https://doi.org/10.1103/RevModPhys.80.885} {\bibfield
  {journal} {\bibinfo  {journal} {Rev. Mod. Phys.}\ }\textbf {\bibinfo {volume}
  {80}},\ \bibinfo {pages} {885} (\bibinfo {year} {2008})}\BibitemShut
  {NoStop}%
\bibitem [{\citenamefont {Norcia}\ and\ \citenamefont
  {Ferlaino}(2021)}]{Norcia2021nof}%
  \BibitemOpen
  \bibfield  {author} {\bibinfo {author} {\bibfnamefont {M.~A.}\ \bibnamefont
  {Norcia}}\ and\ \bibinfo {author} {\bibfnamefont {F.}~\bibnamefont
  {Ferlaino}},\ }\bibfield  {title} {\bibinfo {title} {Developments in atomic
  control using ultracold magnetic lanthanides},\ }\bibfield  {journal}
  {\bibinfo  {journal} {Nature Physics}\ }\href
  {https://doi.org/10.1038/s41567-021-01398-7} {10.1038/s41567-021-01398-7}
  (\bibinfo {year} {2021})\BibitemShut {NoStop}%
\bibitem [{\citenamefont {Chin}\ \emph {et~al.}(2010)\citenamefont {Chin},
  \citenamefont {Grimm}, \citenamefont {Julienne},\ and\ \citenamefont
  {Tiesinga}}]{Chin2010fri}%
  \BibitemOpen
  \bibfield  {author} {\bibinfo {author} {\bibfnamefont {C.}~\bibnamefont
  {Chin}}, \bibinfo {author} {\bibfnamefont {R.}~\bibnamefont {Grimm}},
  \bibinfo {author} {\bibfnamefont {P.}~\bibnamefont {Julienne}},\ and\
  \bibinfo {author} {\bibfnamefont {E.}~\bibnamefont {Tiesinga}},\ }\bibfield
  {title} {\bibinfo {title} {Feshbach resonances in ultracold gases},\ }\href
  {https://doi.org/10.1103/RevModPhys.82.1225} {\bibfield  {journal} {\bibinfo
  {journal} {Rev. Mod. Phys.}\ }\textbf {\bibinfo {volume} {82}},\ \bibinfo
  {pages} {1225} (\bibinfo {year} {2010})}\BibitemShut {NoStop}%
\bibitem [{\citenamefont {Frisch}\ \emph {et~al.}(2014)\citenamefont {Frisch},
  \citenamefont {Mark}, \citenamefont {Aikawa}, \citenamefont {Ferlaino},
  \citenamefont {Bohn}, \citenamefont {Makrides}, \citenamefont {Petrov},\ and\
  \citenamefont {Kotochigova}}]{Frisch2014qci}%
  \BibitemOpen
  \bibfield  {author} {\bibinfo {author} {\bibfnamefont {A.}~\bibnamefont
  {Frisch}}, \bibinfo {author} {\bibfnamefont {M.}~\bibnamefont {Mark}},
  \bibinfo {author} {\bibfnamefont {K.}~\bibnamefont {Aikawa}}, \bibinfo
  {author} {\bibfnamefont {F.}~\bibnamefont {Ferlaino}}, \bibinfo {author}
  {\bibfnamefont {J.~L.}\ \bibnamefont {Bohn}}, \bibinfo {author}
  {\bibfnamefont {C.}~\bibnamefont {Makrides}}, \bibinfo {author}
  {\bibfnamefont {A.}~\bibnamefont {Petrov}},\ and\ \bibinfo {author}
  {\bibfnamefont {S.}~\bibnamefont {Kotochigova}},\ }\bibfield  {title}
  {\bibinfo {title} {Quantum chaos in ultracold collisions of gas-phase erbium
  atoms},\ }\href {https://doi.org/10.1038/nature13137} {\bibfield  {journal}
  {\bibinfo  {journal} {Nature}\ }\textbf {\bibinfo {volume} {507}},\ \bibinfo
  {pages} {475} (\bibinfo {year} {2014})}\BibitemShut {NoStop}%
\bibitem [{\citenamefont {Maier}\ \emph
  {et~al.}(2015{\natexlab{a}})\citenamefont {Maier}, \citenamefont {Kadau},
  \citenamefont {Schmitt}, \citenamefont {Wenzel}, \citenamefont
  {Ferrier-Barbut}, \citenamefont {Pfau}, \citenamefont {Frisch}, \citenamefont
  {Baier}, \citenamefont {Aikawa}, \citenamefont {Chomaz}, \citenamefont
  {Mark}, \citenamefont {Ferlaino}, \citenamefont {Makrides}, \citenamefont
  {Tiesinga}, \citenamefont {Petrov},\ and\ \citenamefont
  {Kotochigova}}]{Maier2015eoc}%
  \BibitemOpen
  \bibfield  {author} {\bibinfo {author} {\bibfnamefont {T.}~\bibnamefont
  {Maier}}, \bibinfo {author} {\bibfnamefont {H.}~\bibnamefont {Kadau}},
  \bibinfo {author} {\bibfnamefont {M.}~\bibnamefont {Schmitt}}, \bibinfo
  {author} {\bibfnamefont {M.}~\bibnamefont {Wenzel}}, \bibinfo {author}
  {\bibfnamefont {I.}~\bibnamefont {Ferrier-Barbut}}, \bibinfo {author}
  {\bibfnamefont {T.}~\bibnamefont {Pfau}}, \bibinfo {author} {\bibfnamefont
  {A.}~\bibnamefont {Frisch}}, \bibinfo {author} {\bibfnamefont
  {S.}~\bibnamefont {Baier}}, \bibinfo {author} {\bibfnamefont
  {K.}~\bibnamefont {Aikawa}}, \bibinfo {author} {\bibfnamefont
  {L.}~\bibnamefont {Chomaz}}, \bibinfo {author} {\bibfnamefont {M.~J.}\
  \bibnamefont {Mark}}, \bibinfo {author} {\bibfnamefont {F.}~\bibnamefont
  {Ferlaino}}, \bibinfo {author} {\bibfnamefont {C.}~\bibnamefont {Makrides}},
  \bibinfo {author} {\bibfnamefont {E.}~\bibnamefont {Tiesinga}}, \bibinfo
  {author} {\bibfnamefont {A.}~\bibnamefont {Petrov}},\ and\ \bibinfo {author}
  {\bibfnamefont {S.}~\bibnamefont {Kotochigova}},\ }\bibfield  {title}
  {\bibinfo {title} {{Emergence of Chaotic Scattering in Ultracold Er and
  Dy}},\ }\href {https://doi.org/10.1103/PhysRevX.5.041029} {\bibfield
  {journal} {\bibinfo  {journal} {Phys. Rev. X}\ }\textbf {\bibinfo {volume}
  {5}},\ \bibinfo {pages} {041029} (\bibinfo {year}
  {2015}{\natexlab{a}})}\BibitemShut {NoStop}%
\bibitem [{\citenamefont {Kotochigova}(2014)}]{Kotochigova2014}%
  \BibitemOpen
  \bibfield  {author} {\bibinfo {author} {\bibfnamefont {S.}~\bibnamefont
  {Kotochigova}},\ }\bibfield  {title} {\bibinfo {title} {{Controlling
  interactions between highly magnetic atoms with Feshbach resonances}},\
  }\href {https://doi.org/10.1088/0034-4885/77/9/093901} {\bibfield  {journal}
  {\bibinfo  {journal} {Reports on Progress in Physics}\ }\textbf {\bibinfo
  {volume} {77}},\ \bibinfo {pages} {093901} (\bibinfo {year}
  {2014})}\BibitemShut {NoStop}%
\bibitem [{\citenamefont {Petrov}\ \emph {et~al.}(2012)\citenamefont {Petrov},
  \citenamefont {Tiesinga},\ and\ \citenamefont {Kotochigova}}]{Petrov2012aif}%
  \BibitemOpen
  \bibfield  {author} {\bibinfo {author} {\bibfnamefont {A.}~\bibnamefont
  {Petrov}}, \bibinfo {author} {\bibfnamefont {E.}~\bibnamefont {Tiesinga}},\
  and\ \bibinfo {author} {\bibfnamefont {S.}~\bibnamefont {Kotochigova}},\
  }\bibfield  {title} {\bibinfo {title} {{Anisotropy-Induced Feshbach
  Resonances in a Quantum Dipolar Gas of Highly Magnetic Atoms}},\ }\href
  {https://doi.org/10.1103/PhysRevLett.109.103002} {\bibfield  {journal}
  {\bibinfo  {journal} {Phys. Rev. Lett.}\ }\textbf {\bibinfo {volume} {109}},\
  \bibinfo {pages} {103002} (\bibinfo {year} {2012})}\BibitemShut {NoStop}%
\bibitem [{\citenamefont {B\"ottcher}\ \emph
  {et~al.}(2019{\natexlab{a}})\citenamefont {B\"ottcher}, \citenamefont
  {Schmidt}, \citenamefont {Wenzel}, \citenamefont {Hertkorn}, \citenamefont
  {Guo}, \citenamefont {Langen},\ and\ \citenamefont {Pfau}}]{Boettcher:2019}%
  \BibitemOpen
  \bibfield  {author} {\bibinfo {author} {\bibfnamefont {F.}~\bibnamefont
  {B\"ottcher}}, \bibinfo {author} {\bibfnamefont {J.-N.}\ \bibnamefont
  {Schmidt}}, \bibinfo {author} {\bibfnamefont {M.}~\bibnamefont {Wenzel}},
  \bibinfo {author} {\bibfnamefont {J.}~\bibnamefont {Hertkorn}}, \bibinfo
  {author} {\bibfnamefont {M.}~\bibnamefont {Guo}}, \bibinfo {author}
  {\bibfnamefont {T.}~\bibnamefont {Langen}},\ and\ \bibinfo {author}
  {\bibfnamefont {T.}~\bibnamefont {Pfau}},\ }\bibfield  {title} {\bibinfo
  {title} {{Transient Supersolid Properties in an Array of Dipolar Quantum
  Droplets}},\ }\href {https://doi.org/10.1103/PhysRevX.9.011051} {\bibfield
  {journal} {\bibinfo  {journal} {Phys. Rev. X}\ }\textbf {\bibinfo {volume}
  {9}},\ \bibinfo {pages} {011051} (\bibinfo {year}
  {2019}{\natexlab{a}})}\BibitemShut {NoStop}%
\bibitem [{\citenamefont {Chomaz}\ \emph {et~al.}(2019)\citenamefont {Chomaz},
  \citenamefont {Petter}, \citenamefont {Ilzh\"ofer}, \citenamefont {Natale},
  \citenamefont {Trautmann}, \citenamefont {Politi}, \citenamefont
  {Durastante}, \citenamefont {van Bijnen}, \citenamefont {Patscheider},
  \citenamefont {Sohmen}, \citenamefont {Mark},\ and\ \citenamefont
  {Ferlaino}}]{Chomaz2019lla}%
  \BibitemOpen
  \bibfield  {author} {\bibinfo {author} {\bibfnamefont {L.}~\bibnamefont
  {Chomaz}}, \bibinfo {author} {\bibfnamefont {D.}~\bibnamefont {Petter}},
  \bibinfo {author} {\bibfnamefont {P.}~\bibnamefont {Ilzh\"ofer}}, \bibinfo
  {author} {\bibfnamefont {G.}~\bibnamefont {Natale}}, \bibinfo {author}
  {\bibfnamefont {A.}~\bibnamefont {Trautmann}}, \bibinfo {author}
  {\bibfnamefont {C.}~\bibnamefont {Politi}}, \bibinfo {author} {\bibfnamefont
  {G.}~\bibnamefont {Durastante}}, \bibinfo {author} {\bibfnamefont {R.~M.~W.}\
  \bibnamefont {van Bijnen}}, \bibinfo {author} {\bibfnamefont
  {A.}~\bibnamefont {Patscheider}}, \bibinfo {author} {\bibfnamefont
  {M.}~\bibnamefont {Sohmen}}, \bibinfo {author} {\bibfnamefont {M.~J.}\
  \bibnamefont {Mark}},\ and\ \bibinfo {author} {\bibfnamefont
  {F.}~\bibnamefont {Ferlaino}},\ }\bibfield  {title} {\bibinfo {title}
  {{Long-Lived and Transient Supersolid Behaviors in Dipolar Quantum Gases}},\
  }\href {https://doi.org/10.1103/PhysRevX.9.021012} {\bibfield  {journal}
  {\bibinfo  {journal} {Phys. Rev. X}\ }\textbf {\bibinfo {volume} {9}},\
  \bibinfo {pages} {021012} (\bibinfo {year} {2019})}\BibitemShut {NoStop}%
\bibitem [{\citenamefont {Tanzi}\ \emph {et~al.}(2019)\citenamefont {Tanzi},
  \citenamefont {Lucioni}, \citenamefont {Fam\`a}, \citenamefont {Catani},
  \citenamefont {Fioretti}, \citenamefont {Gabbanini}, \citenamefont {Bisset},
  \citenamefont {Santos},\ and\ \citenamefont {Modugno}}]{Tanzi2019ooa}%
  \BibitemOpen
  \bibfield  {author} {\bibinfo {author} {\bibfnamefont {L.}~\bibnamefont
  {Tanzi}}, \bibinfo {author} {\bibfnamefont {E.}~\bibnamefont {Lucioni}},
  \bibinfo {author} {\bibfnamefont {F.}~\bibnamefont {Fam\`a}}, \bibinfo
  {author} {\bibfnamefont {J.}~\bibnamefont {Catani}}, \bibinfo {author}
  {\bibfnamefont {A.}~\bibnamefont {Fioretti}}, \bibinfo {author}
  {\bibfnamefont {C.}~\bibnamefont {Gabbanini}}, \bibinfo {author}
  {\bibfnamefont {R.~N.}\ \bibnamefont {Bisset}}, \bibinfo {author}
  {\bibfnamefont {L.}~\bibnamefont {Santos}},\ and\ \bibinfo {author}
  {\bibfnamefont {G.}~\bibnamefont {Modugno}},\ }\bibfield  {title} {\bibinfo
  {title} {{Observation of a Dipolar Quantum Gas with Metastable Supersolid
  Properties}},\ }\href {https://doi.org/10.1103/PhysRevLett.122.130405}
  {\bibfield  {journal} {\bibinfo  {journal} {Phys. Rev. Lett.}\ }\textbf
  {\bibinfo {volume} {122}},\ \bibinfo {pages} {130405} (\bibinfo {year}
  {2019})}\BibitemShut {NoStop}%
\bibitem [{\citenamefont {Lima}\ and\ \citenamefont
  {Pelster}(2012)}]{Lima2012bmf}%
  \BibitemOpen
  \bibfield  {author} {\bibinfo {author} {\bibfnamefont {A.~R.~P.}\
  \bibnamefont {Lima}}\ and\ \bibinfo {author} {\bibfnamefont {A.}~\bibnamefont
  {Pelster}},\ }\bibfield  {title} {\bibinfo {title} {Beyond mean-field
  low-lying excitations of dipolar bose gases},\ }\href
  {https://doi.org/10.1103/PhysRevA.86.063609} {\bibfield  {journal} {\bibinfo
  {journal} {Phys. Rev. A}\ }\textbf {\bibinfo {volume} {86}},\ \bibinfo
  {pages} {063609} (\bibinfo {year} {2012})}\BibitemShut {NoStop}%
\bibitem [{\citenamefont {Chomaz}\ \emph {et~al.}(2018)\citenamefont {Chomaz},
  \citenamefont {van Bijnen}, \citenamefont {Petter}, \citenamefont {Faraoni},
  \citenamefont {Baier}, \citenamefont {Becher}, \citenamefont {Mark},
  \citenamefont {W\"achtler}, \citenamefont {Santos},\ and\ \citenamefont
  {Ferlaino}}]{Chomaz2018oor}%
  \BibitemOpen
  \bibfield  {author} {\bibinfo {author} {\bibfnamefont {L.}~\bibnamefont
  {Chomaz}}, \bibinfo {author} {\bibfnamefont {R.~M.~W.}\ \bibnamefont {van
  Bijnen}}, \bibinfo {author} {\bibfnamefont {D.}~\bibnamefont {Petter}},
  \bibinfo {author} {\bibfnamefont {G.}~\bibnamefont {Faraoni}}, \bibinfo
  {author} {\bibfnamefont {S.}~\bibnamefont {Baier}}, \bibinfo {author}
  {\bibfnamefont {J.~H.}\ \bibnamefont {Becher}}, \bibinfo {author}
  {\bibfnamefont {M.~J.}\ \bibnamefont {Mark}}, \bibinfo {author}
  {\bibfnamefont {F.}~\bibnamefont {W\"achtler}}, \bibinfo {author}
  {\bibfnamefont {L.}~\bibnamefont {Santos}},\ and\ \bibinfo {author}
  {\bibfnamefont {F.}~\bibnamefont {Ferlaino}},\ }\bibfield  {title} {\bibinfo
  {title} {Observation of roton mode population in a dipolar quantum gas},\
  }\href {https://www.nature.com/articles/s41567-018-0054-7} {\bibfield
  {journal} {\bibinfo  {journal} {Nat. Phys.}\ }\textbf {\bibinfo {volume}
  {14}},\ \bibinfo {pages} {442} (\bibinfo {year} {2018})}\BibitemShut
  {NoStop}%
\bibitem [{\citenamefont {Petter}\ \emph {et~al.}(2019)\citenamefont {Petter},
  \citenamefont {Natale}, \citenamefont {van Bijnen}, \citenamefont
  {Patscheider}, \citenamefont {Mark}, \citenamefont {Chomaz},\ and\
  \citenamefont {Ferlaino}}]{Petter2019ptr}%
  \BibitemOpen
  \bibfield  {author} {\bibinfo {author} {\bibfnamefont {D.}~\bibnamefont
  {Petter}}, \bibinfo {author} {\bibfnamefont {G.}~\bibnamefont {Natale}},
  \bibinfo {author} {\bibfnamefont {R.~M.~W.}\ \bibnamefont {van Bijnen}},
  \bibinfo {author} {\bibfnamefont {A.}~\bibnamefont {Patscheider}}, \bibinfo
  {author} {\bibfnamefont {M.~J.}\ \bibnamefont {Mark}}, \bibinfo {author}
  {\bibfnamefont {L.}~\bibnamefont {Chomaz}},\ and\ \bibinfo {author}
  {\bibfnamefont {F.}~\bibnamefont {Ferlaino}},\ }\bibfield  {title} {\bibinfo
  {title} {{Probing the Roton Excitation Spectrum of a Stable Dipolar Bose
  Gas}},\ }\href {https://doi.org/10.1103/PhysRevLett.122.183401} {\bibfield
  {journal} {\bibinfo  {journal} {Phys. Rev. Lett.}\ }\textbf {\bibinfo
  {volume} {122}},\ \bibinfo {pages} {183401} (\bibinfo {year}
  {2019})}\BibitemShut {NoStop}%
\bibitem [{\citenamefont {B\"ottcher}\ \emph
  {et~al.}(2019{\natexlab{b}})\citenamefont {B\"ottcher}, \citenamefont
  {Wenzel}, \citenamefont {Schmidt}, \citenamefont {Guo}, \citenamefont
  {Langen}, \citenamefont {Ferrier-Barbut}, \citenamefont {Pfau}, \citenamefont
  {Bomb\'{\i}n}, \citenamefont {S\'anchez-Baena}, \citenamefont {Boronat},\
  and\ \citenamefont {Mazzanti}}]{Boettcher2019ddq}%
  \BibitemOpen
  \bibfield  {author} {\bibinfo {author} {\bibfnamefont {F.}~\bibnamefont
  {B\"ottcher}}, \bibinfo {author} {\bibfnamefont {M.}~\bibnamefont {Wenzel}},
  \bibinfo {author} {\bibfnamefont {J.-N.}\ \bibnamefont {Schmidt}}, \bibinfo
  {author} {\bibfnamefont {M.}~\bibnamefont {Guo}}, \bibinfo {author}
  {\bibfnamefont {T.}~\bibnamefont {Langen}}, \bibinfo {author} {\bibfnamefont
  {I.}~\bibnamefont {Ferrier-Barbut}}, \bibinfo {author} {\bibfnamefont
  {T.}~\bibnamefont {Pfau}}, \bibinfo {author} {\bibfnamefont {R.}~\bibnamefont
  {Bomb\'{\i}n}}, \bibinfo {author} {\bibfnamefont {J.}~\bibnamefont
  {S\'anchez-Baena}}, \bibinfo {author} {\bibfnamefont {J.}~\bibnamefont
  {Boronat}},\ and\ \bibinfo {author} {\bibfnamefont {F.}~\bibnamefont
  {Mazzanti}},\ }\bibfield  {title} {\bibinfo {title} {{Dilute dipolar quantum
  droplets beyond the extended Gross-Pitaevskii equation}},\ }\href
  {https://doi.org/10.1103/PhysRevResearch.1.033088} {\bibfield  {journal}
  {\bibinfo  {journal} {Phys. Rev. Research}\ }\textbf {\bibinfo {volume}
  {1}},\ \bibinfo {pages} {033088} (\bibinfo {year}
  {2019}{\natexlab{b}})}\BibitemShut {NoStop}%
\bibitem [{\citenamefont {Maier}\ \emph
  {et~al.}(2015{\natexlab{b}})\citenamefont {Maier}, \citenamefont
  {Ferrier-Barbut}, \citenamefont {Kadau}, \citenamefont {Schmitt},
  \citenamefont {Wenzel}, \citenamefont {Wink}, \citenamefont {Pfau},
  \citenamefont {Jachymski},\ and\ \citenamefont {Julienne}}]{Maier2015}%
  \BibitemOpen
  \bibfield  {author} {\bibinfo {author} {\bibfnamefont {T.}~\bibnamefont
  {Maier}}, \bibinfo {author} {\bibfnamefont {I.}~\bibnamefont
  {Ferrier-Barbut}}, \bibinfo {author} {\bibfnamefont {H.}~\bibnamefont
  {Kadau}}, \bibinfo {author} {\bibfnamefont {M.}~\bibnamefont {Schmitt}},
  \bibinfo {author} {\bibfnamefont {M.}~\bibnamefont {Wenzel}}, \bibinfo
  {author} {\bibfnamefont {C.}~\bibnamefont {Wink}}, \bibinfo {author}
  {\bibfnamefont {T.}~\bibnamefont {Pfau}}, \bibinfo {author} {\bibfnamefont
  {K.}~\bibnamefont {Jachymski}},\ and\ \bibinfo {author} {\bibfnamefont
  {P.~S.}\ \bibnamefont {Julienne}},\ }\bibfield  {title} {\bibinfo {title}
  {{Broad universal Feshbach resonances in the chaotic spectrum of dysprosium
  atoms}},\ }\href {https://doi.org/10.1103/PhysRevA.92.060702} {\bibfield
  {journal} {\bibinfo  {journal} {Phys. Rev. A}\ }\textbf {\bibinfo {volume}
  {92}},\ \bibinfo {pages} {060702} (\bibinfo {year}
  {2015}{\natexlab{b}})}\BibitemShut {NoStop}%
\bibitem [{\citenamefont {Lucioni}\ \emph {et~al.}(2018)\citenamefont
  {Lucioni}, \citenamefont {Tanzi}, \citenamefont {Fregosi}, \citenamefont
  {Catani}, \citenamefont {Gozzini}, \citenamefont {Inguscio}, \citenamefont
  {Fioretti}, \citenamefont {Gabbanini},\ and\ \citenamefont
  {Modugno}}]{Lucioni2018}%
  \BibitemOpen
  \bibfield  {author} {\bibinfo {author} {\bibfnamefont {E.}~\bibnamefont
  {Lucioni}}, \bibinfo {author} {\bibfnamefont {L.}~\bibnamefont {Tanzi}},
  \bibinfo {author} {\bibfnamefont {A.}~\bibnamefont {Fregosi}}, \bibinfo
  {author} {\bibfnamefont {J.}~\bibnamefont {Catani}}, \bibinfo {author}
  {\bibfnamefont {S.}~\bibnamefont {Gozzini}}, \bibinfo {author} {\bibfnamefont
  {M.}~\bibnamefont {Inguscio}}, \bibinfo {author} {\bibfnamefont
  {A.}~\bibnamefont {Fioretti}}, \bibinfo {author} {\bibfnamefont
  {C.}~\bibnamefont {Gabbanini}},\ and\ \bibinfo {author} {\bibfnamefont
  {G.}~\bibnamefont {Modugno}},\ }\bibfield  {title} {\bibinfo {title}
  {{Dysprosium dipolar Bose-Einstein condensate with broad Feshbach
  resonances}},\ }\href {https://doi.org/10.1103/PhysRevA.97.060701} {\bibfield
   {journal} {\bibinfo  {journal} {Phys. Rev. A}\ }\textbf {\bibinfo {volume}
  {97}},\ \bibinfo {pages} {060701} (\bibinfo {year} {2018})}\BibitemShut
  {NoStop}%
\bibitem [{\citenamefont {Tang}\ \emph {et~al.}(2016)\citenamefont {Tang},
  \citenamefont {Sykes}, \citenamefont {Burdick}, \citenamefont {DiSciacca},
  \citenamefont {Petrov},\ and\ \citenamefont {Lev}}]{Tang2016}%
  \BibitemOpen
  \bibfield  {author} {\bibinfo {author} {\bibfnamefont {Y.}~\bibnamefont
  {Tang}}, \bibinfo {author} {\bibfnamefont {A.~G.}\ \bibnamefont {Sykes}},
  \bibinfo {author} {\bibfnamefont {N.~Q.}\ \bibnamefont {Burdick}}, \bibinfo
  {author} {\bibfnamefont {J.~M.}\ \bibnamefont {DiSciacca}}, \bibinfo {author}
  {\bibfnamefont {D.~S.}\ \bibnamefont {Petrov}},\ and\ \bibinfo {author}
  {\bibfnamefont {B.~L.}\ \bibnamefont {Lev}},\ }\bibfield  {title} {\bibinfo
  {title} {{Anisotropic Expansion of a Thermal Dipolar Bose Gas}},\ }\href
  {https://doi.org/10.1103/PhysRevLett.117.155301} {\bibfield  {journal}
  {\bibinfo  {journal} {Phys. Rev. Lett.}\ }\textbf {\bibinfo {volume} {117}},\
  \bibinfo {pages} {155301} (\bibinfo {year} {2016})}\BibitemShut {NoStop}%
\bibitem [{\citenamefont {Bohn}\ and\ \citenamefont {Jin}(2014)}]{Jin2014}%
  \BibitemOpen
  \bibfield  {author} {\bibinfo {author} {\bibfnamefont {J.~L.}\ \bibnamefont
  {Bohn}}\ and\ \bibinfo {author} {\bibfnamefont {D.~S.}\ \bibnamefont {Jin}},\
  }\bibfield  {title} {\bibinfo {title} {Differential scattering and
  rethermalization in ultracold dipolar gases},\ }\href
  {https://doi.org/10.1103/PhysRevA.89.022702} {\bibfield  {journal} {\bibinfo
  {journal} {Phys. Rev. A}\ }\textbf {\bibinfo {volume} {89}},\ \bibinfo
  {pages} {022702} (\bibinfo {year} {2014})}\BibitemShut {NoStop}%
\bibitem [{\citenamefont {Aikawa}\ \emph {et~al.}(2014)\citenamefont {Aikawa},
  \citenamefont {Frisch}, \citenamefont {Mark}, \citenamefont {Baier},
  \citenamefont {Grimm}, \citenamefont {Bohn}, \citenamefont {Jin},
  \citenamefont {Bruun},\ and\ \citenamefont {Ferlaino}}]{Aikawa2014ard}%
  \BibitemOpen
  \bibfield  {author} {\bibinfo {author} {\bibfnamefont {K.}~\bibnamefont
  {Aikawa}}, \bibinfo {author} {\bibfnamefont {A.}~\bibnamefont {Frisch}},
  \bibinfo {author} {\bibfnamefont {M.}~\bibnamefont {Mark}}, \bibinfo {author}
  {\bibfnamefont {S.}~\bibnamefont {Baier}}, \bibinfo {author} {\bibfnamefont
  {R.}~\bibnamefont {Grimm}}, \bibinfo {author} {\bibfnamefont {J.~L.}\
  \bibnamefont {Bohn}}, \bibinfo {author} {\bibfnamefont {D.~S.}\ \bibnamefont
  {Jin}}, \bibinfo {author} {\bibfnamefont {G.~M.}\ \bibnamefont {Bruun}},\
  and\ \bibinfo {author} {\bibfnamefont {F.}~\bibnamefont {Ferlaino}},\
  }\bibfield  {title} {\bibinfo {title} {{Anisotropic Relaxation Dynamics in a
  Dipolar Fermi Gas Driven Out of Equilibrium}},\ }\href
  {https://doi.org/10.1103/PhysRevLett.113.263201} {\bibfield  {journal}
  {\bibinfo  {journal} {Phys. Rev. Lett.}\ }\textbf {\bibinfo {volume} {113}},\
  \bibinfo {pages} {263201} (\bibinfo {year} {2014})}\BibitemShut {NoStop}%
\bibitem [{\citenamefont {Sykes}\ and\ \citenamefont {Bohn}(2015)}]{Bohn2015}%
  \BibitemOpen
  \bibfield  {author} {\bibinfo {author} {\bibfnamefont {A.~G.}\ \bibnamefont
  {Sykes}}\ and\ \bibinfo {author} {\bibfnamefont {J.~L.}\ \bibnamefont
  {Bohn}},\ }\bibfield  {title} {\bibinfo {title} {Nonequilibrium dynamics of
  an ultracold dipolar gas},\ }\href
  {https://doi.org/10.1103/PhysRevA.91.013625} {\bibfield  {journal} {\bibinfo
  {journal} {Phys. Rev. A}\ }\textbf {\bibinfo {volume} {91}},\ \bibinfo
  {pages} {013625} (\bibinfo {year} {2015})}\BibitemShut {NoStop}%
\bibitem [{\citenamefont {Tang}\ \emph {et~al.}(2015)\citenamefont {Tang},
  \citenamefont {Sykes}, \citenamefont {Burdick}, \citenamefont {Bohn},\ and\
  \citenamefont {Lev}}]{Tang:2015}%
  \BibitemOpen
  \bibfield  {author} {\bibinfo {author} {\bibfnamefont {Y.}~\bibnamefont
  {Tang}}, \bibinfo {author} {\bibfnamefont {A.}~\bibnamefont {Sykes}},
  \bibinfo {author} {\bibfnamefont {N.~Q.}\ \bibnamefont {Burdick}}, \bibinfo
  {author} {\bibfnamefont {J.~L.}\ \bibnamefont {Bohn}},\ and\ \bibinfo
  {author} {\bibfnamefont {B.~L.}\ \bibnamefont {Lev}},\ }\bibfield  {title}
  {\bibinfo {title} {$s$-wave scattering lengths of the strongly dipolar bosons
  $^{162}\mathrm{Dy}$ and $^{164}\mathrm{Dy}$},\ }\href
  {https://doi.org/10.1103/PhysRevA.92.022703} {\bibfield  {journal} {\bibinfo
  {journal} {Phys. Rev. A}\ }\textbf {\bibinfo {volume} {92}},\ \bibinfo
  {pages} {022703} (\bibinfo {year} {2015})}\BibitemShut {NoStop}%
\bibitem [{\citenamefont {Kollath}\ \emph {et~al.}(2006)\citenamefont
  {Kollath}, \citenamefont {Iucci}, \citenamefont {Giamarchi}, \citenamefont
  {Hofstetter},\ and\ \citenamefont {Schollw\"ock}}]{Kollath:2006}%
  \BibitemOpen
  \bibfield  {author} {\bibinfo {author} {\bibfnamefont {C.}~\bibnamefont
  {Kollath}}, \bibinfo {author} {\bibfnamefont {A.}~\bibnamefont {Iucci}},
  \bibinfo {author} {\bibfnamefont {T.}~\bibnamefont {Giamarchi}}, \bibinfo
  {author} {\bibfnamefont {W.}~\bibnamefont {Hofstetter}},\ and\ \bibinfo
  {author} {\bibfnamefont {U.}~\bibnamefont {Schollw\"ock}},\ }\bibfield
  {title} {\bibinfo {title} {{Spectroscopy of Ultracold Atoms by Periodic
  Lattice Modulations}},\ }\href
  {https://doi.org/10.1103/PhysRevLett.97.050402} {\bibfield  {journal}
  {\bibinfo  {journal} {Phys. Rev. Lett.}\ }\textbf {\bibinfo {volume} {97}},\
  \bibinfo {pages} {050402} (\bibinfo {year} {2006})}\BibitemShut {NoStop}%
\bibitem [{\citenamefont {Baier}\ \emph {et~al.}(2016)\citenamefont {Baier},
  \citenamefont {Mark}, \citenamefont {Petter}, \citenamefont {Aikawa},
  \citenamefont {Chomaz}, \citenamefont {Cai}, \citenamefont {Baranov},
  \citenamefont {Zoller},\ and\ \citenamefont {Ferlaino}}]{Baier2016ebh}%
  \BibitemOpen
  \bibfield  {author} {\bibinfo {author} {\bibfnamefont {S.}~\bibnamefont
  {Baier}}, \bibinfo {author} {\bibfnamefont {M.~J.}\ \bibnamefont {Mark}},
  \bibinfo {author} {\bibfnamefont {D.}~\bibnamefont {Petter}}, \bibinfo
  {author} {\bibfnamefont {K.}~\bibnamefont {Aikawa}}, \bibinfo {author}
  {\bibfnamefont {L.}~\bibnamefont {Chomaz}}, \bibinfo {author} {\bibfnamefont
  {Z.}~\bibnamefont {Cai}}, \bibinfo {author} {\bibfnamefont {M.}~\bibnamefont
  {Baranov}}, \bibinfo {author} {\bibfnamefont {P.}~\bibnamefont {Zoller}},\
  and\ \bibinfo {author} {\bibfnamefont {F.}~\bibnamefont {Ferlaino}},\
  }\bibfield  {title} {\bibinfo {title} {{Extended Bose-Hubbard models with
  ultracold magnetic atoms}},\ }\href {https://doi.org/10.1126/science.aac9812}
  {\bibfield  {journal} {\bibinfo  {journal} {Science}\ }\textbf {\bibinfo
  {volume} {352}},\ \bibinfo {pages} {201} (\bibinfo {year}
  {2016})}\BibitemShut {NoStop}%
\bibitem [{\citenamefont {Yi}\ and\ \citenamefont {You}(2001)}]{Yi2001tco}%
  \BibitemOpen
  \bibfield  {author} {\bibinfo {author} {\bibfnamefont {S.}~\bibnamefont
  {Yi}}\ and\ \bibinfo {author} {\bibfnamefont {L.}~\bibnamefont {You}},\
  }\bibfield  {title} {\bibinfo {title} {Trapped condensates of atoms with
  dipole interactions},\ }\href {https://doi.org/10.1103/PhysRevA.63.053607}
  {\bibfield  {journal} {\bibinfo  {journal} {Phys. Rev. A}\ }\textbf {\bibinfo
  {volume} {63}},\ \bibinfo {pages} {053607} (\bibinfo {year}
  {2001})}\BibitemShut {NoStop}%
\bibitem [{\citenamefont {Ronen}\ \emph {et~al.}(2006)\citenamefont {Ronen},
  \citenamefont {Bortolotti}, \citenamefont {Blume},\ and\ \citenamefont
  {Bohn}}]{Ronen2006dbe}%
  \BibitemOpen
  \bibfield  {author} {\bibinfo {author} {\bibfnamefont {S.}~\bibnamefont
  {Ronen}}, \bibinfo {author} {\bibfnamefont {D.~C.~E.}\ \bibnamefont
  {Bortolotti}}, \bibinfo {author} {\bibfnamefont {D.}~\bibnamefont {Blume}},\
  and\ \bibinfo {author} {\bibfnamefont {J.~L.}\ \bibnamefont {Bohn}},\
  }\bibfield  {title} {\bibinfo {title} {{Dipolar Bose-Einstein condensates
  with dipole-dependent scattering length}},\ }\href
  {https://doi.org/10.1103/PhysRevA.74.033611} {\bibfield  {journal} {\bibinfo
  {journal} {Phys. Rev. A}\ }\textbf {\bibinfo {volume} {74}},\ \bibinfo
  {pages} {033611} (\bibinfo {year} {2006})}\BibitemShut {NoStop}%
\bibitem [{\citenamefont {Bortolotti}\ \emph {et~al.}(2006)\citenamefont
  {Bortolotti}, \citenamefont {Ronen}, \citenamefont {Bohn},\ and\
  \citenamefont {Blume}}]{Bortolotti2006sli}%
  \BibitemOpen
  \bibfield  {author} {\bibinfo {author} {\bibfnamefont {D.~C.~E.}\
  \bibnamefont {Bortolotti}}, \bibinfo {author} {\bibfnamefont
  {S.}~\bibnamefont {Ronen}}, \bibinfo {author} {\bibfnamefont {J.~L.}\
  \bibnamefont {Bohn}},\ and\ \bibinfo {author} {\bibfnamefont
  {D.}~\bibnamefont {Blume}},\ }\bibfield  {title} {\bibinfo {title}
  {{Scattering Length Instability in Dipolar Bose-Einstein Condensates}},\
  }\href {https://doi.org/10.1103/PhysRevLett.97.160402} {\bibfield  {journal}
  {\bibinfo  {journal} {Phys. Rev. Lett.}\ }\textbf {\bibinfo {volume} {97}},\
  \bibinfo {pages} {160402} (\bibinfo {year} {2006})}\BibitemShut {NoStop}%
\bibitem [{\citenamefont {O\l{}dziejewski}\ and\ \citenamefont
  {Jachymski}(2016)}]{Oldziejewski2016pos}%
  \BibitemOpen
  \bibfield  {author} {\bibinfo {author} {\bibfnamefont {R.}~\bibnamefont
  {O\l{}dziejewski}}\ and\ \bibinfo {author} {\bibfnamefont {K.}~\bibnamefont
  {Jachymski}},\ }\bibfield  {title} {\bibinfo {title} {Properties of strongly
  dipolar bose gases beyond the born approximation},\ }\href
  {https://doi.org/10.1103/PhysRevA.94.063638} {\bibfield  {journal} {\bibinfo
  {journal} {Phys. Rev. A}\ }\textbf {\bibinfo {volume} {94}},\ \bibinfo
  {pages} {063638} (\bibinfo {year} {2016})}\BibitemShut {NoStop}%
\bibitem [{\citenamefont {Ferrier-Barbut}\ \emph {et~al.}(2018)\citenamefont
  {Ferrier-Barbut}, \citenamefont {Wenzel}, \citenamefont {B\"ottcher},
  \citenamefont {Langen}, \citenamefont {Isoard}, \citenamefont {Stringari},\
  and\ \citenamefont {Pfau}}]{FerrierBarbut2018smo}%
  \BibitemOpen
  \bibfield  {author} {\bibinfo {author} {\bibfnamefont {I.}~\bibnamefont
  {Ferrier-Barbut}}, \bibinfo {author} {\bibfnamefont {M.}~\bibnamefont
  {Wenzel}}, \bibinfo {author} {\bibfnamefont {F.}~\bibnamefont {B\"ottcher}},
  \bibinfo {author} {\bibfnamefont {T.}~\bibnamefont {Langen}}, \bibinfo
  {author} {\bibfnamefont {M.}~\bibnamefont {Isoard}}, \bibinfo {author}
  {\bibfnamefont {S.}~\bibnamefont {Stringari}},\ and\ \bibinfo {author}
  {\bibfnamefont {T.}~\bibnamefont {Pfau}},\ }\bibfield  {title} {\bibinfo
  {title} {{Scissors Mode of Dipolar Quantum Droplets of Dysprosium Atoms}},\
  }\href {https://doi.org/10.1103/PhysRevLett.120.160402} {\bibfield  {journal}
  {\bibinfo  {journal} {Phys. Rev. Lett.}\ }\textbf {\bibinfo {volume} {120}},\
  \bibinfo {pages} {160402} (\bibinfo {year} {2018})}\BibitemShut {NoStop}%
\bibitem [{\citenamefont {Chomaz}\ \emph {et~al.}(2016)\citenamefont {Chomaz},
  \citenamefont {Baier}, \citenamefont {Petter}, \citenamefont {Mark},
  \citenamefont {W\"achtler}, \citenamefont {Santos},\ and\ \citenamefont
  {Ferlaino}}]{Chomaz2016qfd}%
  \BibitemOpen
  \bibfield  {author} {\bibinfo {author} {\bibfnamefont {L.}~\bibnamefont
  {Chomaz}}, \bibinfo {author} {\bibfnamefont {S.}~\bibnamefont {Baier}},
  \bibinfo {author} {\bibfnamefont {D.}~\bibnamefont {Petter}}, \bibinfo
  {author} {\bibfnamefont {M.~J.}\ \bibnamefont {Mark}}, \bibinfo {author}
  {\bibfnamefont {F.}~\bibnamefont {W\"achtler}}, \bibinfo {author}
  {\bibfnamefont {L.}~\bibnamefont {Santos}},\ and\ \bibinfo {author}
  {\bibfnamefont {F.}~\bibnamefont {Ferlaino}},\ }\bibfield  {title} {\bibinfo
  {title} {{Quantum-Fluctuation-Driven Crossover from a Dilute Bose-Einstein
  Condensate to a Macrodroplet in a Dipolar Quantum Fluid}},\ }\href
  {https://doi.org/10.1103/PhysRevX.6.041039} {\bibfield  {journal} {\bibinfo
  {journal} {Phys. Rev. X}\ }\textbf {\bibinfo {volume} {6}},\ \bibinfo {pages}
  {041039} (\bibinfo {year} {2016})}\BibitemShut {NoStop}%
\bibitem [{\citenamefont {Monroe}\ \emph {et~al.}(1993)\citenamefont {Monroe},
  \citenamefont {Cornell}, \citenamefont {Sackett}, \citenamefont {Myatt},\
  and\ \citenamefont {Wieman}}]{Monroe1993}%
  \BibitemOpen
  \bibfield  {author} {\bibinfo {author} {\bibfnamefont {C.~R.}\ \bibnamefont
  {Monroe}}, \bibinfo {author} {\bibfnamefont {E.~A.}\ \bibnamefont {Cornell}},
  \bibinfo {author} {\bibfnamefont {C.~A.}\ \bibnamefont {Sackett}}, \bibinfo
  {author} {\bibfnamefont {C.~J.}\ \bibnamefont {Myatt}},\ and\ \bibinfo
  {author} {\bibfnamefont {C.~E.}\ \bibnamefont {Wieman}},\ }\bibfield  {title}
  {\bibinfo {title} {{Measurement of Cs-Cs elastic scattering at T=30
  \ensuremath{\mu}K}},\ }\href {https://doi.org/10.1103/PhysRevLett.70.414}
  {\bibfield  {journal} {\bibinfo  {journal} {Phys. Rev. Lett.}\ }\textbf
  {\bibinfo {volume} {70}},\ \bibinfo {pages} {414} (\bibinfo {year}
  {1993})}\BibitemShut {NoStop}%
\bibitem [{\citenamefont {Newbury}\ \emph {et~al.}(1995)\citenamefont
  {Newbury}, \citenamefont {Myatt},\ and\ \citenamefont
  {Wieman}}]{Newbury1995sec}%
  \BibitemOpen
  \bibfield  {author} {\bibinfo {author} {\bibfnamefont {N.~R.}\ \bibnamefont
  {Newbury}}, \bibinfo {author} {\bibfnamefont {C.~J.}\ \bibnamefont {Myatt}},\
  and\ \bibinfo {author} {\bibfnamefont {C.~E.}\ \bibnamefont {Wieman}},\
  }\bibfield  {title} {\bibinfo {title} {s-wave elastic collisions between cold
  ground-state $^{87}\mathrm{Rb}$ atoms},\ }\href
  {https://doi.org/10.1103/PhysRevA.51.R2680} {\bibfield  {journal} {\bibinfo
  {journal} {Phys. Rev. A}\ }\textbf {\bibinfo {volume} {51}},\ \bibinfo
  {pages} {R2680} (\bibinfo {year} {1995})}\BibitemShut {NoStop}%
\bibitem [{\citenamefont {Davis}\ \emph {et~al.}(1995)\citenamefont {Davis},
  \citenamefont {Mewes}, \citenamefont {Joffe}, \citenamefont {Andrews},\ and\
  \citenamefont {Ketterle}}]{Davis1995eco}%
  \BibitemOpen
  \bibfield  {author} {\bibinfo {author} {\bibfnamefont {K.~B.}\ \bibnamefont
  {Davis}}, \bibinfo {author} {\bibfnamefont {M.-O.}\ \bibnamefont {Mewes}},
  \bibinfo {author} {\bibfnamefont {M.~A.}\ \bibnamefont {Joffe}}, \bibinfo
  {author} {\bibfnamefont {M.~R.}\ \bibnamefont {Andrews}},\ and\ \bibinfo
  {author} {\bibfnamefont {W.}~\bibnamefont {Ketterle}},\ }\bibfield  {title}
  {\bibinfo {title} {{Evaporative Cooling of Sodium Atoms}},\ }\href
  {https://doi.org/10.1103/PhysRevLett.74.5202} {\bibfield  {journal} {\bibinfo
   {journal} {Phys. Rev. Lett.}\ }\textbf {\bibinfo {volume} {74}},\ \bibinfo
  {pages} {5202} (\bibinfo {year} {1995})}\BibitemShut {NoStop}%
\bibitem [{\citenamefont {Hopkins}\ \emph {et~al.}(2000)\citenamefont
  {Hopkins}, \citenamefont {Webster}, \citenamefont {Arlt}, \citenamefont
  {Bance}, \citenamefont {Cornish}, \citenamefont {Marag\`o},\ and\
  \citenamefont {Foot}}]{Hopkins2000moe}%
  \BibitemOpen
  \bibfield  {author} {\bibinfo {author} {\bibfnamefont {S.~A.}\ \bibnamefont
  {Hopkins}}, \bibinfo {author} {\bibfnamefont {S.}~\bibnamefont {Webster}},
  \bibinfo {author} {\bibfnamefont {J.}~\bibnamefont {Arlt}}, \bibinfo {author}
  {\bibfnamefont {P.}~\bibnamefont {Bance}}, \bibinfo {author} {\bibfnamefont
  {S.}~\bibnamefont {Cornish}}, \bibinfo {author} {\bibfnamefont
  {O.}~\bibnamefont {Marag\`o}},\ and\ \bibinfo {author} {\bibfnamefont
  {C.~J.}\ \bibnamefont {Foot}},\ }\bibfield  {title} {\bibinfo {title}
  {Measurement of elastic cross section for cold cesium collisions},\ }\href
  {https://doi.org/10.1103/PhysRevA.61.032707} {\bibfield  {journal} {\bibinfo
  {journal} {Phys. Rev. A}\ }\textbf {\bibinfo {volume} {61}},\ \bibinfo
  {pages} {032707} (\bibinfo {year} {2000})}\BibitemShut {NoStop}%
\bibitem [{\citenamefont {Schmidt}\ \emph {et~al.}(2003)\citenamefont
  {Schmidt}, \citenamefont {Hensler}, \citenamefont {Werner}, \citenamefont
  {Griesmaier}, \citenamefont {G\"orlitz}, \citenamefont {Pfau},\ and\
  \citenamefont {Simoni}}]{Schmidt2003}%
  \BibitemOpen
  \bibfield  {author} {\bibinfo {author} {\bibfnamefont {P.~O.}\ \bibnamefont
  {Schmidt}}, \bibinfo {author} {\bibfnamefont {S.}~\bibnamefont {Hensler}},
  \bibinfo {author} {\bibfnamefont {J.}~\bibnamefont {Werner}}, \bibinfo
  {author} {\bibfnamefont {A.}~\bibnamefont {Griesmaier}}, \bibinfo {author}
  {\bibfnamefont {A.}~\bibnamefont {G\"orlitz}}, \bibinfo {author}
  {\bibfnamefont {T.}~\bibnamefont {Pfau}},\ and\ \bibinfo {author}
  {\bibfnamefont {A.}~\bibnamefont {Simoni}},\ }\bibfield  {title} {\bibinfo
  {title} {{Determination of the $s$-Wave Scattering Length of Chromium}},\
  }\href {https://doi.org/10.1103/PhysRevLett.91.193201} {\bibfield  {journal}
  {\bibinfo  {journal} {Phys. Rev. Lett.}\ }\textbf {\bibinfo {volume} {91}},\
  \bibinfo {pages} {193201} (\bibinfo {year} {2003})}\BibitemShut {NoStop}%
\bibitem [{\citenamefont {Valtolina}\ \emph {et~al.}(2020)\citenamefont
  {Valtolina}, \citenamefont {Matsuda}, \citenamefont {Tobias}, \citenamefont
  {Li}, \citenamefont {De~Marco},\ and\ \citenamefont {Ye}}]{Valtolina2020deo}%
  \BibitemOpen
  \bibfield  {author} {\bibinfo {author} {\bibfnamefont {G.}~\bibnamefont
  {Valtolina}}, \bibinfo {author} {\bibfnamefont {K.}~\bibnamefont {Matsuda}},
  \bibinfo {author} {\bibfnamefont {W.~G.}\ \bibnamefont {Tobias}}, \bibinfo
  {author} {\bibfnamefont {J.-R.}\ \bibnamefont {Li}}, \bibinfo {author}
  {\bibfnamefont {L.}~\bibnamefont {De~Marco}},\ and\ \bibinfo {author}
  {\bibfnamefont {J.}~\bibnamefont {Ye}},\ }\bibfield  {title} {\bibinfo
  {title} {{Dipolar evaporation of reactive molecules to below the Fermi
  temperature}},\ }\href {https://doi.org/10.1038/s41586-020-2980-7} {\bibfield
   {journal} {\bibinfo  {journal} {Nature}\ }\textbf {\bibinfo {volume}
  {588}},\ \bibinfo {pages} {239} (\bibinfo {year} {2020})}\BibitemShut
  {NoStop}%
\bibitem [{\citenamefont {Li}\ \emph {et~al.}(2021)\citenamefont {Li},
  \citenamefont {Tobias}, \citenamefont {Matsuda}, \citenamefont {Miller},
  \citenamefont {Valtolina}, \citenamefont {De~Marco}, \citenamefont {Wang},
  \citenamefont {Lassabli{\`e}re}, \citenamefont {Qu{\'e}m{\'e}ner},
  \citenamefont {Bohn},\ and\ \citenamefont {Ye}}]{Li2021tof}%
  \BibitemOpen
  \bibfield  {author} {\bibinfo {author} {\bibfnamefont {J.-R.}\ \bibnamefont
  {Li}}, \bibinfo {author} {\bibfnamefont {W.~G.}\ \bibnamefont {Tobias}},
  \bibinfo {author} {\bibfnamefont {K.}~\bibnamefont {Matsuda}}, \bibinfo
  {author} {\bibfnamefont {C.}~\bibnamefont {Miller}}, \bibinfo {author}
  {\bibfnamefont {G.}~\bibnamefont {Valtolina}}, \bibinfo {author}
  {\bibfnamefont {L.}~\bibnamefont {De~Marco}}, \bibinfo {author}
  {\bibfnamefont {R.~R.~W.}\ \bibnamefont {Wang}}, \bibinfo {author}
  {\bibfnamefont {L.}~\bibnamefont {Lassabli{\`e}re}}, \bibinfo {author}
  {\bibfnamefont {G.}~\bibnamefont {Qu{\'e}m{\'e}ner}}, \bibinfo {author}
  {\bibfnamefont {J.~L.}\ \bibnamefont {Bohn}},\ and\ \bibinfo {author}
  {\bibfnamefont {J.}~\bibnamefont {Ye}},\ }\bibfield  {title} {\bibinfo
  {title} {Tuning of dipolar interactions and evaporative cooling in a
  three-dimensional molecular quantum gas},\ }\href
  {https://doi.org/10.1038/s41567-021-01329-6} {\bibfield  {journal} {\bibinfo
  {journal} {Nature Physics}\ }\textbf {\bibinfo {volume} {17}},\ \bibinfo
  {pages} {1144} (\bibinfo {year} {2021})}\BibitemShut {NoStop}%
\bibitem [{\citenamefont {Wang}\ and\ \citenamefont {Bohn}(2021)}]{Wang2021}%
  \BibitemOpen
  \bibfield  {author} {\bibinfo {author} {\bibfnamefont {R.~R.~W.}\
  \bibnamefont {Wang}}\ and\ \bibinfo {author} {\bibfnamefont {J.~L.}\
  \bibnamefont {Bohn}},\ }\bibfield  {title} {\bibinfo {title} {Anisotropic
  thermalization of dilute dipolar gases},\ }\href
  {https://doi.org/10.1103/PhysRevA.103.063320} {\bibfield  {journal} {\bibinfo
   {journal} {Phys. Rev. A}\ }\textbf {\bibinfo {volume} {103}},\ \bibinfo
  {pages} {063320} (\bibinfo {year} {2021})}\BibitemShut {NoStop}%
\bibitem [{\citenamefont {Aikawa}\ \emph {et~al.}(2012)\citenamefont {Aikawa},
  \citenamefont {Frisch}, \citenamefont {Mark}, \citenamefont {Baier},
  \citenamefont {Rietzler}, \citenamefont {Grimm},\ and\ \citenamefont
  {Ferlaino}}]{Aikawa2012bec}%
  \BibitemOpen
  \bibfield  {author} {\bibinfo {author} {\bibfnamefont {K.}~\bibnamefont
  {Aikawa}}, \bibinfo {author} {\bibfnamefont {A.}~\bibnamefont {Frisch}},
  \bibinfo {author} {\bibfnamefont {M.}~\bibnamefont {Mark}}, \bibinfo {author}
  {\bibfnamefont {S.}~\bibnamefont {Baier}}, \bibinfo {author} {\bibfnamefont
  {A.}~\bibnamefont {Rietzler}}, \bibinfo {author} {\bibfnamefont
  {R.}~\bibnamefont {Grimm}},\ and\ \bibinfo {author} {\bibfnamefont
  {F.}~\bibnamefont {Ferlaino}},\ }\bibfield  {title} {\bibinfo {title}
  {{Bose-Einstein Condensation of Erbium}},\ }\href
  {https://doi.org/10.1103/PhysRevLett.108.210401} {\bibfield  {journal}
  {\bibinfo  {journal} {Phys. Rev. Lett.}\ }\textbf {\bibinfo {volume} {108}},\
  \bibinfo {pages} {210401} (\bibinfo {year} {2012})}\BibitemShut {NoStop}%
\bibitem [{\citenamefont {Frisch}\ \emph {et~al.}(2012)\citenamefont {Frisch},
  \citenamefont {Aikawa}, \citenamefont {Mark}, \citenamefont {Rietzler},
  \citenamefont {Schindler}, \citenamefont {Zupani\ifmmode~\check{c}\else
  \v{c}\fi{}}, \citenamefont {Grimm},\ and\ \citenamefont
  {Ferlaino}}]{Frisch2012nlm}%
  \BibitemOpen
  \bibfield  {author} {\bibinfo {author} {\bibfnamefont {A.}~\bibnamefont
  {Frisch}}, \bibinfo {author} {\bibfnamefont {K.}~\bibnamefont {Aikawa}},
  \bibinfo {author} {\bibfnamefont {M.}~\bibnamefont {Mark}}, \bibinfo {author}
  {\bibfnamefont {A.}~\bibnamefont {Rietzler}}, \bibinfo {author}
  {\bibfnamefont {J.}~\bibnamefont {Schindler}}, \bibinfo {author}
  {\bibfnamefont {E.}~\bibnamefont {Zupani\ifmmode~\check{c}\else \v{c}\fi{}}},
  \bibinfo {author} {\bibfnamefont {R.}~\bibnamefont {Grimm}},\ and\ \bibinfo
  {author} {\bibfnamefont {F.}~\bibnamefont {Ferlaino}},\ }\bibfield  {title}
  {\bibinfo {title} {Narrow-line magneto-optical trap for erbium},\ }\href
  {https://doi.org/10.1103/PhysRevA.85.051401} {\bibfield  {journal} {\bibinfo
  {journal} {Phys. Rev. A}\ }\textbf {\bibinfo {volume} {85}},\ \bibinfo
  {pages} {051401} (\bibinfo {year} {2012})}\BibitemShut {NoStop}%
\bibitem [{\citenamefont {Reif}(1965)}]{Reif1965}%
  \BibitemOpen
  \bibfield  {author} {\bibinfo {author} {\bibfnamefont {F.}~\bibnamefont
  {Reif}},\ }\href@noop {} {\emph {\bibinfo {title} {Fundamentals of
  statistical and thermal physics}}}\ (\bibinfo  {publisher} {McGraw-Hill},\
  \bibinfo {year} {1965})\BibitemShut {NoStop}%
\bibitem [{\citenamefont {Wang}\ \emph {et~al.}(2020)\citenamefont {Wang},
  \citenamefont {Sykes},\ and\ \citenamefont {Bohn}}]{Wang2020}%
  \BibitemOpen
  \bibfield  {author} {\bibinfo {author} {\bibfnamefont {R.~R.~W.}\
  \bibnamefont {Wang}}, \bibinfo {author} {\bibfnamefont {A.~G.}\ \bibnamefont
  {Sykes}},\ and\ \bibinfo {author} {\bibfnamefont {J.~L.}\ \bibnamefont
  {Bohn}},\ }\bibfield  {title} {\bibinfo {title} {Linear response of a
  periodically driven thermal dipolar gas},\ }\href
  {https://doi.org/10.1103/PhysRevA.102.033336} {\bibfield  {journal} {\bibinfo
   {journal} {Phys. Rev. A}\ }\textbf {\bibinfo {volume} {102}},\ \bibinfo
  {pages} {033336} (\bibinfo {year} {2020})}\BibitemShut {NoStop}%
\bibitem [{\citenamefont {DeMarco}\ \emph {et~al.}(1999)\citenamefont
  {DeMarco}, \citenamefont {Bohn}, \citenamefont {Burke}, \citenamefont
  {Holland},\ and\ \citenamefont {Jin}}]{DeMarco1999}%
  \BibitemOpen
  \bibfield  {author} {\bibinfo {author} {\bibfnamefont {B.}~\bibnamefont
  {DeMarco}}, \bibinfo {author} {\bibfnamefont {J.~L.}\ \bibnamefont {Bohn}},
  \bibinfo {author} {\bibfnamefont {J.~P.}\ \bibnamefont {Burke}}, \bibinfo
  {author} {\bibfnamefont {M.}~\bibnamefont {Holland}},\ and\ \bibinfo {author}
  {\bibfnamefont {D.~S.}\ \bibnamefont {Jin}},\ }\bibfield  {title} {\bibinfo
  {title} {{Measurement of $\mathit{p}$-Wave Threshold Law Using Evaporatively
  Cooled Fermionic Atoms}},\ }\href
  {https://doi.org/10.1103/PhysRevLett.82.4208} {\bibfield  {journal} {\bibinfo
   {journal} {Phys. Rev. Lett.}\ }\textbf {\bibinfo {volume} {82}},\ \bibinfo
  {pages} {4208} (\bibinfo {year} {1999})}\BibitemShut {NoStop}%
\bibitem [{sup()}]{supmat}%
  \BibitemOpen
  \href@noop {} {}\bibinfo {note} {See Supplemental Material at [URL], which
  includes details on the theory calculations and data analysis.}\BibitemShut
  {Stop}%
\bibitem [{\citenamefont {Baier}\ \emph {et~al.}(2018)\citenamefont {Baier},
  \citenamefont {Petter}, \citenamefont {Becher}, \citenamefont {Patscheider},
  \citenamefont {Natale}, \citenamefont {Chomaz}, \citenamefont {Mark},\ and\
  \citenamefont {Ferlaino}}]{Baier2018roa}%
  \BibitemOpen
  \bibfield  {author} {\bibinfo {author} {\bibfnamefont {S.}~\bibnamefont
  {Baier}}, \bibinfo {author} {\bibfnamefont {D.}~\bibnamefont {Petter}},
  \bibinfo {author} {\bibfnamefont {J.~H.}\ \bibnamefont {Becher}}, \bibinfo
  {author} {\bibfnamefont {A.}~\bibnamefont {Patscheider}}, \bibinfo {author}
  {\bibfnamefont {G.}~\bibnamefont {Natale}}, \bibinfo {author} {\bibfnamefont
  {L.}~\bibnamefont {Chomaz}}, \bibinfo {author} {\bibfnamefont {M.~J.}\
  \bibnamefont {Mark}},\ and\ \bibinfo {author} {\bibfnamefont
  {F.}~\bibnamefont {Ferlaino}},\ }\bibfield  {title} {\bibinfo {title}
  {{Realization of a Strongly Interacting Fermi Gas of Dipolar Atoms}},\ }\href
  {https://doi.org/10.1103/PhysRevLett.121.093602} {\bibfield  {journal}
  {\bibinfo  {journal} {Phys. Rev. Lett.}\ }\textbf {\bibinfo {volume} {121}},\
  \bibinfo {pages} {093602} (\bibinfo {year} {2018})}\BibitemShut {NoStop}%
\bibitem [{\citenamefont {Greiner}\ \emph {et~al.}(2002)\citenamefont
  {Greiner}, \citenamefont {Mandel}, \citenamefont {Esslinger}, \citenamefont
  {H{\"a}nsch},\ and\ \citenamefont {Bloch}}]{Greiner2002qpt}%
  \BibitemOpen
  \bibfield  {author} {\bibinfo {author} {\bibfnamefont {M.}~\bibnamefont
  {Greiner}}, \bibinfo {author} {\bibfnamefont {O.}~\bibnamefont {Mandel}},
  \bibinfo {author} {\bibfnamefont {T.}~\bibnamefont {Esslinger}}, \bibinfo
  {author} {\bibfnamefont {T.~W.}\ \bibnamefont {H{\"a}nsch}},\ and\ \bibinfo
  {author} {\bibfnamefont {I.}~\bibnamefont {Bloch}},\ }\bibfield  {title}
  {\bibinfo {title} {{Quantum phase transition from a superfluid to a Mott
  insulator in a gas of ultracold atoms}},\ }\href
  {https://doi.org/10.1038/415039a} {\bibfield  {journal} {\bibinfo  {journal}
  {Nature}\ }\textbf {\bibinfo {volume} {415}},\ \bibinfo {pages} {39}
  (\bibinfo {year} {2002})}\BibitemShut {NoStop}%
\bibitem [{\citenamefont {Jachymski}\ and\ \citenamefont
  {Julienne}(2013)}]{Jachymski2013}%
  \BibitemOpen
  \bibfield  {author} {\bibinfo {author} {\bibfnamefont {K.}~\bibnamefont
  {Jachymski}}\ and\ \bibinfo {author} {\bibfnamefont {P.~S.}\ \bibnamefont
  {Julienne}},\ }\bibfield  {title} {\bibinfo {title} {{Analytical model of
  overlapping Feshbach resonances}},\ }\href
  {https://doi.org/10.1103/PhysRevA.88.052701} {\bibfield  {journal} {\bibinfo
  {journal} {Phys. Rev. A}\ }\textbf {\bibinfo {volume} {88}},\ \bibinfo
  {pages} {052701} (\bibinfo {year} {2013})}\BibitemShut {NoStop}%
\bibitem [{\citenamefont {Kraemer}\ \emph {et~al.}(2006)\citenamefont
  {Kraemer}, \citenamefont {Mark}, \citenamefont {Waldburger}, \citenamefont
  {Danzl}, \citenamefont {Chin}, \citenamefont {Engeser}, \citenamefont
  {Lange}, \citenamefont {Pilch}, \citenamefont {Jaakkola}, \citenamefont
  {N{\"a}gerl},\ and\ \citenamefont {Grimm}}]{Kraemer2006efe}%
  \BibitemOpen
  \bibfield  {author} {\bibinfo {author} {\bibfnamefont {T.}~\bibnamefont
  {Kraemer}}, \bibinfo {author} {\bibfnamefont {M.}~\bibnamefont {Mark}},
  \bibinfo {author} {\bibfnamefont {P.}~\bibnamefont {Waldburger}}, \bibinfo
  {author} {\bibfnamefont {J.~G.}\ \bibnamefont {Danzl}}, \bibinfo {author}
  {\bibfnamefont {C.}~\bibnamefont {Chin}}, \bibinfo {author} {\bibfnamefont
  {B.}~\bibnamefont {Engeser}}, \bibinfo {author} {\bibfnamefont {A.~D.}\
  \bibnamefont {Lange}}, \bibinfo {author} {\bibfnamefont {K.}~\bibnamefont
  {Pilch}}, \bibinfo {author} {\bibfnamefont {A.}~\bibnamefont {Jaakkola}},
  \bibinfo {author} {\bibfnamefont {H.-C.}\ \bibnamefont {N{\"a}gerl}},\ and\
  \bibinfo {author} {\bibfnamefont {R.}~\bibnamefont {Grimm}},\ }\bibfield
  {title} {\bibinfo {title} {{Evidence for Efimov quantum states in an
  ultracold gas of caesium atoms}},\ }\href
  {https://doi.org/10.1038/nature04626} {\bibfield  {journal} {\bibinfo
  {journal} {Nature}\ }\textbf {\bibinfo {volume} {440}},\ \bibinfo {pages}
  {315} (\bibinfo {year} {2006})}\BibitemShut {NoStop}%
\bibitem [{\citenamefont {Deb}\ and\ \citenamefont {You}(2001)}]{Deb2001}%
  \BibitemOpen
  \bibfield  {author} {\bibinfo {author} {\bibfnamefont {B.}~\bibnamefont
  {Deb}}\ and\ \bibinfo {author} {\bibfnamefont {L.}~\bibnamefont {You}},\
  }\bibfield  {title} {\bibinfo {title} {Low-energy atomic collision with
  dipole interactions},\ }\href {https://doi.org/10.1103/PhysRevA.64.022717}
  {\bibfield  {journal} {\bibinfo  {journal} {Phys. Rev. A}\ }\textbf {\bibinfo
  {volume} {64}},\ \bibinfo {pages} {022717} (\bibinfo {year}
  {2001})}\BibitemShut {NoStop}%
\bibitem [{\citenamefont {Gribakin}\ and\ \citenamefont
  {Flambaum}(1993)}]{Gribakin1993}%
  \BibitemOpen
  \bibfield  {author} {\bibinfo {author} {\bibfnamefont {G.~F.}\ \bibnamefont
  {Gribakin}}\ and\ \bibinfo {author} {\bibfnamefont {V.~V.}\ \bibnamefont
  {Flambaum}},\ }\bibfield  {title} {\bibinfo {title} {Calculation of the
  scattering length in atomic collisions using the semiclassical
  approximation},\ }\href {https://doi.org/10.1103/PhysRevA.48.546} {\bibfield
  {journal} {\bibinfo  {journal} {Phys. Rev. A}\ }\textbf {\bibinfo {volume}
  {48}},\ \bibinfo {pages} {546} (\bibinfo {year} {1993})}\BibitemShut
  {NoStop}%
\bibitem [{\citenamefont {Kitagawa}\ \emph {et~al.}(2008)\citenamefont
  {Kitagawa}, \citenamefont {Enomoto}, \citenamefont {Kasa}, \citenamefont
  {Takahashi}, \citenamefont {Ciury\l{}o}, \citenamefont {Naidon},\ and\
  \citenamefont {Julienne}}]{Kitagawa2008}%
  \BibitemOpen
  \bibfield  {author} {\bibinfo {author} {\bibfnamefont {M.}~\bibnamefont
  {Kitagawa}}, \bibinfo {author} {\bibfnamefont {K.}~\bibnamefont {Enomoto}},
  \bibinfo {author} {\bibfnamefont {K.}~\bibnamefont {Kasa}}, \bibinfo {author}
  {\bibfnamefont {Y.}~\bibnamefont {Takahashi}}, \bibinfo {author}
  {\bibfnamefont {R.}~\bibnamefont {Ciury\l{}o}}, \bibinfo {author}
  {\bibfnamefont {P.}~\bibnamefont {Naidon}},\ and\ \bibinfo {author}
  {\bibfnamefont {P.~S.}\ \bibnamefont {Julienne}},\ }\bibfield  {title}
  {\bibinfo {title} {Two-color photoassociation spectroscopy of ytterbium atoms
  and the precise determinations of $s$-wave scattering lengths},\ }\href
  {https://doi.org/10.1103/PhysRevA.77.012719} {\bibfield  {journal} {\bibinfo
  {journal} {Phys. Rev. A}\ }\textbf {\bibinfo {volume} {77}},\ \bibinfo
  {pages} {012719} (\bibinfo {year} {2008})}\BibitemShut {NoStop}%
\bibitem [{\citenamefont {Borkowski}\ \emph {et~al.}(2013)\citenamefont
  {Borkowski}, \citenamefont {\ifmmode~\dot{Z}\else \.{Z}\fi{}uchowski},
  \citenamefont {Ciury\l{}o}, \citenamefont {Julienne}, \citenamefont
  {Kedziera}, \citenamefont {Mentel}, \citenamefont {Tecmer}, \citenamefont
  {M\"unchow}, \citenamefont {Bruni},\ and\ \citenamefont
  {G\"orlitz}}]{Borkowski2013sli}%
  \BibitemOpen
  \bibfield  {author} {\bibinfo {author} {\bibfnamefont {M.}~\bibnamefont
  {Borkowski}}, \bibinfo {author} {\bibfnamefont {P.~S.}\ \bibnamefont
  {\ifmmode~\dot{Z}\else \.{Z}\fi{}uchowski}}, \bibinfo {author} {\bibfnamefont
  {R.}~\bibnamefont {Ciury\l{}o}}, \bibinfo {author} {\bibfnamefont {P.~S.}\
  \bibnamefont {Julienne}}, \bibinfo {author} {\bibfnamefont {D.}~\bibnamefont
  {Kedziera}}, \bibinfo {author} {\bibfnamefont {L.}~\bibnamefont {Mentel}},
  \bibinfo {author} {\bibfnamefont {P.}~\bibnamefont {Tecmer}}, \bibinfo
  {author} {\bibfnamefont {F.}~\bibnamefont {M\"unchow}}, \bibinfo {author}
  {\bibfnamefont {C.}~\bibnamefont {Bruni}},\ and\ \bibinfo {author}
  {\bibfnamefont {A.}~\bibnamefont {G\"orlitz}},\ }\bibfield  {title} {\bibinfo
  {title} {{Scattering lengths in isotopologues of the RbYb system}},\ }\href
  {https://doi.org/10.1103/PhysRevA.88.052708} {\bibfield  {journal} {\bibinfo
  {journal} {Phys. Rev. A}\ }\textbf {\bibinfo {volume} {88}},\ \bibinfo
  {pages} {052708} (\bibinfo {year} {2013})}\BibitemShut {NoStop}%
\bibitem [{\citenamefont {Bird}(2013)}]{Bird13_CSIPP}%
  \BibitemOpen
  \bibfield  {author} {\bibinfo {author} {\bibfnamefont {G.~A.}\ \bibnamefont
  {Bird}},\ }\href@noop {} {\emph {\bibinfo {title} {The DSMC method}}}\
  (\bibinfo  {publisher} {CreateSpace Independent Publishing Platform},\
  \bibinfo {year} {2013})\BibitemShut {NoStop}%
\bibitem [{\citenamefont {Durastante}\ \emph {et~al.}(2020)\citenamefont
  {Durastante}, \citenamefont {Politi}, \citenamefont {Sohmen}, \citenamefont
  {Ilzh\"ofer}, \citenamefont {Mark}, \citenamefont {Norcia},\ and\
  \citenamefont {Ferlaino}}]{Durastante2020fri}%
  \BibitemOpen
  \bibfield  {author} {\bibinfo {author} {\bibfnamefont {G.}~\bibnamefont
  {Durastante}}, \bibinfo {author} {\bibfnamefont {C.}~\bibnamefont {Politi}},
  \bibinfo {author} {\bibfnamefont {M.}~\bibnamefont {Sohmen}}, \bibinfo
  {author} {\bibfnamefont {P.}~\bibnamefont {Ilzh\"ofer}}, \bibinfo {author}
  {\bibfnamefont {M.~J.}\ \bibnamefont {Mark}}, \bibinfo {author}
  {\bibfnamefont {M.~A.}\ \bibnamefont {Norcia}},\ and\ \bibinfo {author}
  {\bibfnamefont {F.}~\bibnamefont {Ferlaino}},\ }\bibfield  {title} {\bibinfo
  {title} {Feshbach resonances in an erbium-dysprosium dipolar mixture},\
  }\href {https://doi.org/10.1103/PhysRevA.102.033330} {\bibfield  {journal}
  {\bibinfo  {journal} {Phys. Rev. A}\ }\textbf {\bibinfo {volume} {102}},\
  \bibinfo {pages} {033330} (\bibinfo {year} {2020})}\BibitemShut {NoStop}%
\bibitem [{\citenamefont {Ni}(1979)}]{Ni1979tlt}%
  \BibitemOpen
  \bibfield  {author} {\bibinfo {author} {\bibfnamefont {G.-j.}\ \bibnamefont
  {Ni}},\ }\bibfield  {title} {\bibinfo {title} {{The Levinson theorem and its
  generalization in relativistic quantum mechanics}},\ }\href@noop {}
  {\bibfield  {journal} {\bibinfo  {journal} {Chinese Physics C}\ }\textbf
  {\bibinfo {volume} {3}},\ \bibinfo {pages} {432} (\bibinfo {year}
  {1979})}\BibitemShut {NoStop}%
\end{thebibliography}%

%%%%%%%%%%%%%%%%%%%%%%%%%%%%%%%%%%%%%%%%%%%%%%%%%%%%
\clearpage
%\appendix

\renewcommand\thefigure{ S\arabic{figure}}   
\setcounter{figure}{0}   

\section*{Supplementary material}

\subsection{Analytic number of collisions per re-thermalization}

Analytic expressions for $\alpha (\as,\theta)$ can be derived under a short-time approximation, with the Enskog equations
\begin{subequations}
\label{eq:Enskog_eqns}
\begin{align}
    & \dfrac{d \langle q_j^2 \rangle}{d t} - \dfrac{2}{m}\langle q_j p_j \rangle = 0,  \\
    & \dfrac{d \langle p_j^2 \rangle}{d t} + 2m\omega_j^2\langle q_j p_j \rangle = \mathcal{C}[ p_j^2 ], \\ % \Delta p_j^2
    & \dfrac{d \langle q_j p_j \rangle}{dt} - \dfrac{1}{m}\left\langle p_j^2 \right\rangle + m\omega_j^2 \langle q_j^2 \rangle = 0, 
\end{align}
\end{subequations}
where ${r}_j$ and ${p}_j$ are positions and momenta respectively ($j = x, y, z$), and $\mathcal{C}$ is the collision integral. The derivation follows from Ref.~\cite{Wang2021}, but we present a brief outline here for completeness. The gas is assumed close-to-equilibrium, allowing us to treat ${r}_j$ and ${p}_j$ as Gaussian distributed. Thermalization trajectories are then tracked using the Gaussian widths along each axis, to compute the energy differential
\begin{align}
    \langle \chi_j \rangle &\equiv E_j - k_B T_f,
\end{align}
where $T_f = (T_x + T_y + T_z) / 3$ is the final equilibration temperature (obtained from the equipartition theorem), $\langle \ldots \rangle$ denotes an ensemble average assuming a Gaussian phase space distribution whose widths are allowed to vary, and $E_j = { \langle p_j^2 \rangle / (2m) } + m \omega_j^2 \langle r_j^2 \rangle / 2$ is the sum of kinetic and potential energies in the $j$-th direction. The Enskog equations dictate that the relaxation of $\langle \chi_j \rangle$ follows the differential equation
\begin{align}
    \frac{ d \langle \chi_j \rangle }{ dt } = {\cal C}[ \chi_j ].
\end{align}
% where ${\cal C}$ is the collision integral with analytic forms found in Ref.~\cite{Wang2021}. 

For small deviations from equilibrium and at short times, re-thermalization can be approximated with a single decay rate $\gamma$, such that ${\cal C}[ \chi] \approx - \gamma \langle \chi \rangle$. This results in the relation
\begin{align}
    \frac{ d E_j }{dt} = -\gamma_{yj} \left( E_j - k_B T_f \right) = {\cal C}[ E_j ], \label{eq:gamma_yj}
\end{align}
% With $N_{\text{coll}} = n \sigma_{\text{el}} v / \gamma$, we arrive at a concise expression:
% \begin{align}
%     N_{\text{coll}}^x(\theta) = \frac{56}{33 - 17 \cos (4 \theta )}.
% \end{align}
where the subscript on $\gamma_{yj}$ indicates that the gas was excited along $y$, and re-thermalization measured along $j$. This then permits us to compute
\begin{align}
    \alpha_{yj} = \frac{ \bar{n} \bar{\sigma} v_r }{ \gamma_{yj} } = \left( \frac{ E_j - k_B T_f  }{ {\cal C}[ E_j ] } \right) \bar{n} \bar{\sigma} v_r,
\end{align}
which for $j = z$, has the form in Eq.~(\ref{eq:analytic_alpha}). 

\subsection{Fitting Enskog equations to experimental data}

The extraction of the scattering lengths $a_s$, from cross-dimensional thermalization data was done here by means of full numerical solutions to the Enskog equations. To do so, $a_s$ was left as a float parameter in the theory, then varied until a best fit between the theory and experimental data was obtained. A feature we noticed during fitting was the high sensitivity of thermalization rates to variations in the trapping frequencies $\boldsymbol{\omega}$, over the finite-time quench. Measurement uncertainties therefore motivate us to also leave $\boldsymbol{\omega}$ a float parameter, with allowed values within its 1-sigma errorbars. This is applied both to the trapping frequencies before and after the quench.
% $\boldsymbol{\omega}$, before and after the quench, to also be left float parameters within error-bars. 

We performed fits using a $\chi^2$ optimization criterion
\begin{align}
    \min_{ \boldsymbol{\omega}, a_s } \sum_{t = t_0}^{t_\mathrm{end}} \left( \frac{ {T}(t) - {\cal T}_\mathrm{E}\left[ {T}( 0 ); \boldsymbol{\omega}, a_s \right](t) }{ \delta {T}(t) } \right)^2,
\end{align}
where the sum runs over measurement time instances $t$, $T(t)$ is the temperature data from the experiment, $\delta {T}(t)$ is the temperature measurement uncertainty, and ${\cal T}_\mathrm{E}\left[ {T}( 0 ); \boldsymbol{\omega}, a_s \right]$ is the solution to the Enskog equations with initial condition ${T}(0)$, and fit parameters $a_s$ and $\boldsymbol{\omega}$.

To reduce biasing of the fits, we run an iterative algorithm that recursively fits $\boldsymbol{\omega}$ and $a_s$ in succession until they converge to stable values. 
% The iterative procedure above requires multiple trials, that 
Such a procedure would take exceedingly long times ($\sim$ weeks) with full Monte Carlo (MC) simulations, but can be done in minutes with the Enskog equations on a current-day
computing device. 
% To add to this, we are also interested in obtaining $a_s$ for multiple magnetic field values (ranging from $B = 0.2$ -- $5$), in several different isotopes of Er ($^{164}$Er, $^{166}$Er and $^{170}$Er). 
% Solutions to the Enskog equations are verified with full Monte Carlo simulations at the end of each iterative cycle. 

\begin{figure}
    \centering
    \includegraphics[width = 0.48\textwidth]{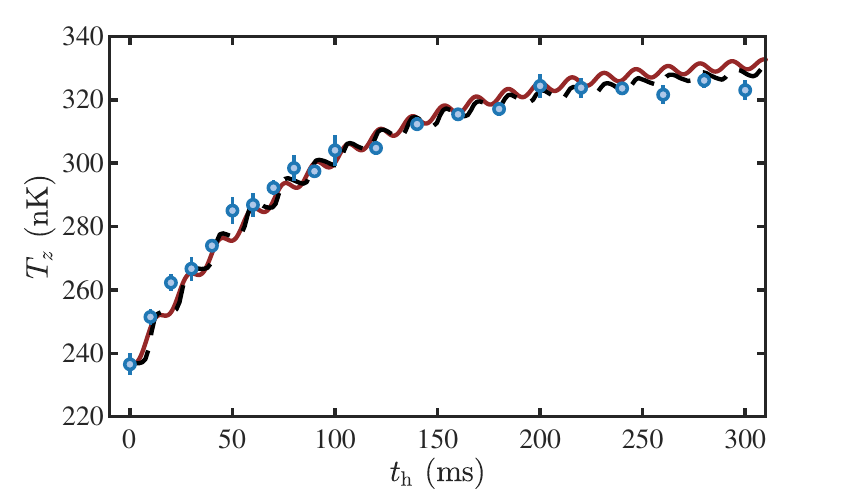}
    \caption{Benchmarking of the Enskog simulation results for $T_z$ (red solid line) with Monte-Carlo simulations (black dashed line). The dataset is the same as in Fig.\,\ref{fig:Figure_CT_and_System}.}
    \label{fig:Figure_Enskog_MC_v1}
\end{figure}

Solutions to the Enskog equations have shown themselves accurate when compared to MC simulations \cite{Wang2020, Wang2021}. We show their accuracy here yet again, using the parameters from the current experimental set-up. An illustrative example is provided in the plot of Fig.~\ref{fig:Figure_Enskog_MC_v1}, comparing an instance of the Enskog solutions (red solid line), MC simulations (black dashed line) and the experimental data (blue circles). 

\subsection{Monte-Carlo simulations including trap anharmonicities}

Optical dipole traps are, in many studies, assumed to be well modeled by purely harmonic potentials.
% \begin{align}
%     U_h(\boldsymbol{r}) = \frac{ 1 }{ 2 } m \sum_j \omega_j^2 q_j^2,
% \end{align}
% where $\omega_j$ are the effectively harmonic trap frequencies and $j = 1,2,3$. 
This may however be inadequate in regimes with significant trap anharmonicity effects, which we currently attribute the density dependence of $\alpha$ to. 
% In such cases, an accurate model of the ODT potential must be adopted for a rigorous theoretical study.
In such cases, the potential is better modeled as two cross-propagating Gaussian-profile beams along the $y$ and $z$ axes (with gravity). This produces the confinement potential % \cite{Baier12_thesis} 
\begin{align}
    V_\mathrm{ODT}(\boldsymbol{r}) &= - \frac{ 2 \tilde{U}_1 P_1 }{ \pi w_{1,x}(z) w_{1,y}(z) } e^{ -2 \left( \frac{x^2}{w_{1,x}^2(z)} + \frac{y^2}{w_{1,y}^2(z)} \right) } \nonumber\\
    &\quad -  \frac{ 2 \tilde{U}_2 P_2 }{ \pi w_{2,x}(y) w_{2,y}(y) }  e^{ -2 \left( \frac{x^2}{w_{2,x}^2(y)} + \frac{z^2}{w_{2,z}^2(y)} \right) } \nonumber\\
    &\quad + m g z, \label{eq:ODT_potential}
\end{align} 
where $P$ is the laser power  $\tilde{U}$ is an atomic polarizability parameter and
\begin{align}
    w(z) = w_{0} \sqrt{ 1 + \frac{ z^2 }{ z_{R}^2} },
\end{align}
with $z_{R}$ and $w_0$ denoting Rayleigh lengths and beam widths respectively.

Such a potential limits the applicability of the aforementioned Enskog equations as formulated in Ref.~\cite{Wang2020}. Instead, more robust MD methods are required to accurately predict thermalization trajectories. We implement a MD simulation similar to that in Ref.~\cite{Bohn2015}, which evolves simulation particles under the action of $V_\mathrm{ODT}$ via the Verlet symplectic integrator
\begin{subequations} \label{eq:Verlet_integration}
\begin{align}
    & \boldsymbol{q}_k = \boldsymbol{r}_k(t) + \frac{\Delta t}{2m}\boldsymbol{p}_k(t), \\
    & \boldsymbol{p}_k(t + \Delta t) = \boldsymbol{p}_k(t) + \boldsymbol{F}_k\Delta t, \\
    & \boldsymbol{r}_k(t + \Delta t) = \boldsymbol{q}_k + \frac{\Delta t}{2m}\boldsymbol{p}_k(t + \Delta t),
\end{align}
\end{subequations}
where subscripts $k$ denote the $k$-th simulation particle, $\Delta t$ is the simulation time-step, $t$ is the time and 
\begin{align}
    % \boldsymbol{F}_k = -\grad ( U_\mathrm{trap} + U_\mathrm{dd} )
    \boldsymbol{F}_k = -\nabla V_\mathrm{ODT}(\boldsymbol{r}_k).
\end{align}
Dipolar collisions are then computed with the direct simulation Monte Carlo method \cite{Bird13_CSIPP}, that determines post-collision momenta via stochastic sampling of the differential cross section. 
% to solve the Boltzmann equation with traps of the form in Eq.~\ref{eq:ODT_potential} . The ODT confinement is applied to the simulated particles as a force
% \begin{align}
%     \boldsymbol{F}_k = -\grad ( U_\mathrm{trap} + U_\mathrm{dd} )
%     \boldsymbol{F}_\mathrm{ODT} = -\nabla V_\mathrm{ODT},
% \end{align}
% which non-trivially couples coordinates axes unlike in harmonic confinement.  

In a preliminary study of the density dependence, ideal Gaussian beam profiles are assumed, along with perfectly accurate beam widths and Rayleigh lengths. Following a trap quench, thermalization of the out-of-equilibrium gas in $V_\mathrm{ODT}$ indeed shows an apparent increase of $\alpha$ with density, qualitatively similar to that observed in the experiment. This effect is absent in simulations with an ideal harmonic trap. Furthermore, in higher density regimes, the simulations with $V_\mathrm{ODT}$ predict the experimentally observed equilibration temperatures more accurately compared to the harmonic trap case. These early findings on density dependence from trap anharmonicities are intriguing, and a cautionary tale for future experiments. However, we do not develop this idea further here and leave such analysis for future works.

\subsection{Fano-Feshbach spectroscopy}

To identify the positions of the Fano-Feshbach resonances we perform high-resolution loss spectroscopy in a cylindrically symmetric trap. We evaporatively cool the atoms until they reach a temperature between $T = \SI{300}{nK}$ and \SI{400}{nK}. At this stage, the atom number is between \num{6e4} and \num{1.2e4} with typical trap frequencies of $(\omega_x, \omega_y, \omega_z) = 2\pi\times (300, 30, 300)\,\si{Hz}$. The exact values depend on the isotope choice. After reaching thermal equilibrium, we change $B$, oriented along the $z$ axis, in \SI{1}{ms} to the desired value and wait for a holding time between \SI{250}{ms} and \SI{500}{ms}. We use different holding times for different datasets to avoid saturation effects of the resonances for higher densities. After the holding time, we measure the atom number using absorption imaging after a time of flight expansion of \SI{25}{ms}. The results of the loss-spectroscopy measurements are shown in Fig.\,\ref{fig:as_166}, Fig.\,\ref{fig:as_164_170} and Fig.\,\ref{fig:as_168}.

\subsection{Scattering length for $^{168}$Er}

\begin{figure*}
    \centering
    \includegraphics[scale =1]{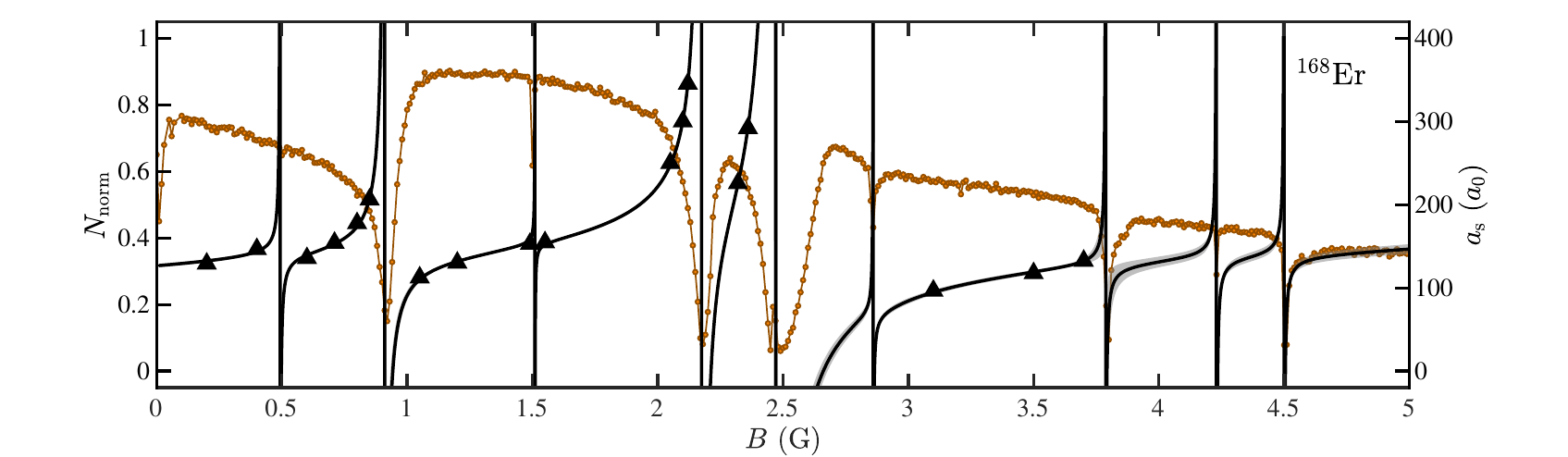}
    \caption{Normalized atom number (yellow circles) and measured scattering lengths $\as$~obtained for $^{168}$Er from lattice modulation spectroscopy measurements. The solid black line represents a fit to \asLMS. The shaded area as well as the error bars denote the standard error.}
    \label{fig:as_168}
\end{figure*}

To obtain $\as$ for the $^{168}$Er isotope, we follow a similar approach as for $^{166}$Er. First, we perform loss spectroscopy to identify the position of Fano-Feshbach resonances. We then transfer the atoms into an optical lattice with a depth of $(s_x, s_y, s_z) = (20, 20, 40) E_{\text{rec}}$ and apply the lattice modulation spectroscopy technique to extract $\as$. The lattice modulation spectroscopy follows the same lines as for the $^{166}$Er isotope; see main text. Fig.\,\ref{fig:as_168} summarizes the results for $^{168}$Er and shows the Fano-Feshbach spectroscopy result as well as $\as$ as a function of $B$ in the magnetic field range from \SI{0}{G} to \SI{5}{G}.

% \subsection{Lattice modulation spectroscopy}

% In brief, we measure the particle-hole excitation gap of a Mott insulator state confined in a deep 3D optical lattice. The energy gap is given by the onsite interaction parameter $U = U_c + U_{dd}$, which is the sum of the contact interaction $U_c$ and the dipolar interaction $U_{dd}$. In the deep lattice approximation, the knowledge of the lattice depths enables us to estimate $U_{dd}$ as well as to specify the relation between $U_c$ and $\as$. Thus, $\as$~is extracted from the measurement of the excitation gap. For the measurement of $U$, the atoms are loaded into a three-dimensional optical lattice created by two retro-reflected laser beams at \SI{532}{nm} in the horizontal plane and by one retro-reflected laser beam at \SI{1064}{nm} along the vertical direction, defined by gravity. The power of the lattice beams is ramped up exponentially in \SI{150}{ms} to reach a lattice depth of $(s_x, s_y, s_z) = (20, 20, 100)$, in units of $E_{\text{rec}} = \SI{4.2}{kHz}
% ~(\SI{1.05}{kHz})$ for \SI{532}{nm} (\SI{1064}{nm}). We create particle-hole-excitations by modulating the power of the lattice beams in the horizontal plane sinusoidally, with a modulation peak-to-peak amplitude of \SI{30}{\%}, for \SI{90}{ms} and measure the recovered BEC fraction after melting of the lattice. At the resonance condition, where the modulation frequency corresponds to the particle hole excitation gap a reduction in the BEC fraction can be observed. 

\subsection{Extracting background scattering length}

To obtain a value for \asbg~, we fit Eq.\,\ref{Eq:a_background_fit} either to $\as$ obtained from the full Enksog equations ($^{164}$Er and $^{170}$Er) or \asLMS~($^{166}$Er and $^{168}$Er). Due to the different numbers of Fano-Feshbach resonances compared to the number of available data points for $\as$, we slightly vary the fitting approach for the individual isotopes. Depending on the position and the width of the resonance, for some resonances, we fix the position $B_i$ to the minimum of the loss feature and keep only the width $\Delta B_i$ as a floating parameter. For the very narrow resonances, which have a negligible influence on the overall scattering behavior, we fix both $B_i$ and $\Delta B_i$. 

Table~\ref{tab:ScatteringlengthSlope} gives the results for the background scattering lengths \asbg~and the slopes $s$ for all four isotopes. Moreover, Tables~\ref{tab:Fit_Results_164}--\ref{tab:Fit_Results_170} contain a detailed listing of all Fano-Feshbach resonances and how they are included in the fitting procedure. Note that for $^{170}$Er, we are aware of the existence of a particularly broad resonance at $\SI{6.91}{G}$~\cite{Durastante2020fri}, which we include with variable width. When looking closely, the onset of this resonance can actually be seen as a reduction of $N$ towards higher magnetic field values in the loss spectroscopy (see Fig.\,\ref{fig:as_164_170}(b)). 

\begin{table}
\caption{Values for \asbg~and $s$ obtained from the fit of Eq.\,\ref{Eq:a_background_fit} to $\as$ for the four bosonic isotopes. The error denotes the fit error of one standard deviation.}
\label{tab:ScatteringlengthSlope}
\begin{ruledtabular}
    \begin{tabular}{ccccc}
        & isotope & \asbg~(\si{\bohr}) & $s$ (\si{\bohr \per G}) & \\
        \hline
        & 164 & \num{52(6)} & \num{9(3)} &  \\
        & 166 & \num{61(3)} & \num{5.4(9)} &  \\
        & 168 & \num{110(2)} & \num{11(2)} &  \\
        & 170 & \num{129(9)} & \num{20(10)} &  \\
    \end{tabular}
\end{ruledtabular}
\end{table}

\begin{table}
\caption{Parameters for the Fano-Feshbach resonances included into the fit of Eq.\,\ref{Eq:a_background_fit} to $\as$ for $^{164}$Er. The error denotes the fit error of one standard deviation. Values without error are fixed in the fitting procedure.}
\label{tab:Fit_Results_164}
\begin{ruledtabular}
    \begin{tabular}{cccc}
        & Position $B_i$ (G) & Width $\Delta B_i$ (G) & \\
        \hline
        & \num{1.52} & \num{0.22(3)}  & \\
        & \num{2.67} & \num{0.005}  & \\
        & \num{2.83} & \num{0.005}  & \\
        & \num{3.26} &  \num{0.10(3)}  & \\
    \end{tabular}
\end{ruledtabular}
\end{table}

\begin{table}
\caption{Parameters for the Fano-Feshbach resonances included into the fit of Eq.\,\ref{Eq:a_background_fit} to $\as$ for $^{166}$Er. The error denotes the fit error of one standard deviation. Values without error are fixed in the fitting procedure.}
\label{tab:Fit_Results_166}
\begin{ruledtabular}
    \begin{tabular}{cccc}
        & Position $B_i$ (G) & Width $\Delta B_i$ (G) & \\
        \hline
        & \num{0.02(5)} & \num{0.05(2)}  & \\
        & \num{3.04(5)} & \num{0.15(2)}  & \\
        & \num{4.208} & \num{0.01}  & \\
        & \num{4.96} & \num{0.005}  & \\
    \end{tabular}
\end{ruledtabular}
\end{table}

\begin{table}
\caption{Parameters for the Fano-Feshbach resonances included into the fit of Eq.\,\ref{Eq:a_background_fit} to $\as$ for $^{168}$Er. The error denotes the fit error of one standard deviation. Values without error are fixed in the fitting procedure.}
\label{tab:Fit_Results_168}
\begin{ruledtabular}
    \begin{tabular}{cccc}
        & Position $B_i$ (G) & Width $\Delta B_i$ (G) & \\
        \hline
        & \num{0.49} & \num{0.005}  & \\
        & \num{0.911(6)} & \num{0.032(2)}  & \\
        & \num{1.51} & \num{0.01}  & \\
        & \num{2.174(4)} & \num{0.038(2)}  & \\
        & \num{2.471(9)} & \num{0.19(1)}  & \\
        & \num{2.86} & \num{0.005}  & \\
        & \num{3.79} & \num{0.006(5)}  & \\
        & \num{4.23} & \num{0.005}  & \\
        & \num{4.5} & \num{0.005}  & \\
    \end{tabular}
\end{ruledtabular}
\end{table}

\begin{table}
\caption{Parameters for the Fano-Feshbach resonances included into the fit of Eq.\,\ref{Eq:a_background_fit} to $\as$ for $^{170}$Er. The error denotes the fit error of one standard deviation. Values without error are fixed in the fitting procedure.}
\label{tab:Fit_Results_170}
\begin{ruledtabular}
    \begin{tabular}{cccc}
        & Position $B_i$ (G) & Width $\Delta B_i$ (G) &  \\
        \hline
        & \num{0.35} & \num{0.005}  & \\
        & \num{0.86} & \num{0.028(12)}  & \\
        & \num{1.12} & \num{0.005}  & \\
        & \num{1.62} & \num{0.01}  & \\
        & \num{2.17} &  \num{0.067(7)}  & \\
        & \num{2.74} & \num{0.134(9)}  & \\
        & \num{3.3} & \num{0.01(1)}  & \\
        & \num{3.57} & \num{0.01}  & \\
        & \num{4.38} & \num{0.005}  & \\
        & \num{4.49} & \num{0.01}  & \\
        & \num{6.91} & \num{0.8(7)}  & \\
    \end{tabular}
\end{ruledtabular}
\end{table}

\subsection{$\chi^2$ analysis for mass scaling}

% \begin{figure}
%     \centering
%     \includegraphics[width = 0.48\textwidth]{Figure_MassScaling_chi2_v1.pdf}
%     \caption{$\chi^2$ as a function of $\phi$ (blue solid line). The dashed line denotes a quadratic fit to the local minima (red stars). The dotted lines indicate the confidence interval of $\Delta \phi = \num{\pm 2.2}$.}
%     \label{fig:Figure_MassScaling_chi2}
% \end{figure}
In this section, we describe our analysis of the background $\as$ of the 4 Er isotopes (Fig.\,\ref{fig:mass_scaling}) with Eq.\,\ref{Eq:a_background_fit}. To find the best fitting parameter $\phi$, we analyze the agreement of the theoretical model in Eq.\,\ref{Eq:a_s_scaling_full_form} with our experimental data. For each value of $\phi$, we calculate the $\chi^2$ via
\begin{equation}
    \chi^2 = \sum_{i=1}^4 \left(\frac{a_s^{\mathrm{mod}}-a_s^i}{\sigma_s^i}\right)^2.
\end{equation}
Here, $a_s^{\mathrm{mod}}$ is the scattering length given by the model for the corresponding $\phi$ and $a_s^i$ and $\sigma_s^i$ are the measured $\as$ with the corresponding standard error. 

The behavior of $\chi^2$ is non-monotonic with the appearance of several minima. We identify the absolute minimum of $\chi^2$ for $\phi = \num{144.03}$. To further obtain an estimate for the error of $\phi$ we fit a quadratic function to the local minima. We extract the limits of the confidence interval by considering the region where $\chi^2 \leq \chi^2 + 1$.

%\begin{figure}
%    \centering
%    \includegraphics[width =0.48\textwidth]{figScalingDensity.pdf}
%    \caption{(a) Experimentally extracted value of $\alpha$ that is necessary to reproduce the LMS results via CDT (blue circles) compared to the value of $\alpha$ that is extracted from the analytic formula (red squares). The solid (dashed) line is a linear fit to the experimental (theoretical) value of $\alpha$ assuming a linear behavior in this region of $\as$. (b) The Figure shows $\alpha$ for different set of measurements where we vary either $\bar{\omega}$, $T$, or $N$. Different symbols and color represent different set of measurements. The solid red line represents the theoretical expectation for $\alpha$.}
%    \label{fig:alpha_scaling_density}
%\end{figure}

\subsection{Hard-core potential for mass scaling}

The model contains the assumption, that the s-wave scattering length is given at large distances by the van-der-Waals potential scaling with $U_\text{vdW}(r)  \propto - C_6/r^6$, with $C_6$ being the Van der Waals coefficient, and at short distances $r < r_c$ by a hard core potential~\cite{Gribakin1993}. In this specific case, the scaling of \asbg~can be described by
\begin{equation}
\label{eq:mass_scaling}
    a_\mathrm{s}^\mathrm{bg} = \Bar{a}\tan(\Phi),
\end{equation}
where $\Bar{a} = \frac{\Gamma (3/4)}{2 \sqrt(2) \Gamma (5/4)} a_c$ with $a_c=\left(\frac{2 m_r C_6}{\hbar^2}\right)^{1/4}$ being the characteristic scattering length scale of the potential, and $\Phi = \frac{a_c^2}{2r_c^2} - \frac{3 \pi}{8}$ is the semi-classical phase~\cite{Gribakin1993}. 

From theoretical calculations in Ref.\,\cite{Kotochigova2014} we use $C_6 = 1723\,$a.u.~ and we estimate from the theoretical interaction potential given in Ref.\cite{Kotochigova2014} that $r_c \approx 4 - \SI{8}{a_0}$. We fit Eq.\,\ref{eq:mass_scaling} to \asbg~of the four bosonic isotopes. Due to a large number of possible local minima, we combine the fitting with a minimization of the $\chi^2$-value while varying the start parameter for $r_c$. We obtain the best agreement for $r_c = \SI{5.05(5)}{\bohr}$. 

In addition, the Levinson theorem~\cite{Ni1979tlt} allows us to estimate the number of bound states $N_B$, which can be calculated from the semi-classical phase $\Phi$ using
\begin{equation}
    N_B = \left[ \frac{\Phi}{\pi} - \frac{5}{8} \right] + 1 ,
\end{equation}
where the square brackets mean the integer part. For the current fitting we obtain $N_B$ ranging from $141$ to $144$, in agreement with the approach in the main text.
We want to emphasize, that this modelling of \asbg~is a simple approach and a more thorough analysis could add deeper valuable insights.

\end{document}